\documentclass[journal,]{IEEEtran}
\usepackage{amsmath,amsfonts}
\usepackage{algorithmic}
\usepackage{algorithm}
\usepackage{array}
\usepackage{textcomp}
\usepackage{stfloats}
\usepackage{url}
\usepackage{verbatim}
\usepackage{graphicx}
\usepackage{cite}
\usepackage{multirow}
\usepackage{bm}
\usepackage{amsfonts}
\usepackage{arydshln}
\usepackage{epstopdf}
\usepackage{amssymb}
\usepackage{amsmath}   
\usepackage{lipsum}
\usepackage{float}
\newcolumntype{L}{>{\raggedright\arraybackslash}X}
\usepackage{ltablex}
\usepackage{siunitx}
\usepackage{booktabs}
\usepackage{gensymb}
\usepackage{mathtools}
\usepackage[caption=false]{subfig}
\newcommand*{\mydprime}{^{\prime\prime}\mkern-1.2mu}
\usepackage{rotating}
\usepackage{tabularx}
\usepackage{multicol}
\usepackage{lipsum}
\usepackage[T1]{fontenc}

\begin{document}

\title{HybridCVLNet: A Hybrid CSI Feedback System and its Domain Adaptation}

\author{Haozhen Li,~\IEEEmembership{Graduate Student Member, IEEE},~Xinyu Gu,~\IEEEmembership{Member, IEEE},~Boyuan Zhang,~Dongliang Li,\\~Zhenyu Liu,~\IEEEmembership{Member, IEEE}~Lin Zhang,~\IEEEmembership{Member, IEEE}
\thanks{This work has been submitted to the IEEE for possible publication.
Copyright may be transferred without notice, after which this version may
no longer be accessible.}
\thanks{H. Li, B. Zhang, D. Li, Z. Liu, L. Zhang are with the School of Artificial Intelligence, Beijing University of Posts and Telecommunications, Beijing,
100876, China (e-mail: \{lihaozhen, zhangboyuan, rhyme\_lee, lzyu, zhanglin\}@bupt.edu.cn).

X. Gu is with the School of Artificial Intelligence, Beijing University of Posts and Telecommunications, Beijing,
100876, China. She is also with the Purple Mountain Laboratories, Nanjing 211111, China (e-mail: guxinyu@bupt.edu.cn).}
}



\maketitle

\begin{abstract}
\textbf{Deep Learning (DL)-based channel state information (CSI) feedback is a promising technique for the transmitter to accurately acquire the CSI of massive multiple-input multiple-output (MIMO) systems. As critical concerns about DL-based physical layer applications, the intra-domain generalizability affected by dataset bias and inter-domain robustness in data drift remain challenging. Therefore, we build on a Hybrid Complex-Valued Lightweight framework, namely the HybridCVLNet, capable of overcoming the dataset bias with regularized hybrid structure and codeword. Meanwhile, a corresponding transductive-based hybrid domain adaptation scheme is proposed to tackle the inter-domain data drift. The experiment verifies that HybridCVLNet achieves stable generalizability and performance gain over the state-of-the-art (SOTA) feedback schemes in an intra-domain heterogeneous dataset. In addition, its transductive-based hybrid domain adaptation scheme is more efficient and superior to the inductive-based transfer learning methods under two inter-domain online re-optimization settings.}
\end{abstract}


\section{Introduction}
\IEEEPARstart{T}{he} advanced massive multiple-input and multiple-output (MIMO) systems leverage precise wireless channel state information (CSI) to yield spatially multiplexing gain and capacity increment. In a frequency division duplexing (FDD) system, the reciprocity of uplink and downlink channels is limited due to the different frequency selective parameters caused by the significant carrier frequency gap. Therefore, the user equipments (UEs) would have to explicitly report the downlink CSI to the serving base station (BS) through an uplink feedback channel. However, the feedback overhead increases linearly as the scale of the transmitting antenna develops. Moreover, as the very-large-scale antenna arrays are becoming promising, the excessive feedback payload and high degrees of freedom of the downlink channel matrices may further undermine the effectiveness of the CSI feedback process.   

Inspired by the image deep compression systems \cite{DeepCompress1}, \cite{DeepCompress2}, the DL-based CSI feedback studies have sprung up \cite{Overview_AICSI} and been highlighted as a study case in the industry \cite{AICSI_in_Industry} as well. A preliminary remarkable deep compression system for CSI feedback called CsiNet in \cite{CsiNet} effectively learns from training samples and is defined as an autoencoder (AE) architecture with end-to-end regression training objectives. Unlike the traditional compressive sensing (CS) approaches, deep compression approximates the optimal mapping between the CSI and a predefined-length compressed latent and feeds into the corresponding recovery model in a trainable manner. 

Generally, existing CSI feedback studies aim at finding a particular mapping that depends on the deployed network and the dataset \emph{domain} (i.e., its input and output space and associated distribution) \cite{Overview_AICSI}. However, we noticed that when an intra-domain heterogeneous dataset appears in the CSI feedback system, the phenomena of inferior generalizability, for instance, the loss gap increasing between training and validation, occur on account of the \emph{dataset bias} (i.e., specific patterns of a heterogeneous dataset are overweighted or overrepresented) \cite{Sensing}. Moreover, when inter-domain unseen data comes, the phenomena of poor robustness, such as empirical inference degradation, arise due to the \emph{data drift} (i.e., the input data distribution varies over time) \cite{Multi-task}. Considering the fact that the downlink channel characteristics by UE mobility, allocated resources, and channel environment, the dataset bias and data drift are ubiquitous, motivating the research on the generalizability and robustness of the CSI feedback system.

Conventional, the generalizability and robustness of neural networks have been proven to degrade as the scale of the network increase while elevating with the enlarging of the visible training dataset. Moreover, the frontiers study of deep learning has deployed over-parameterization systems to boost the inference generalizability while avoiding over-smoothing \cite{Generalize}. Nevertheless, under the required bandwidth-limitation and resource-constrained devices, little research on the generalizability of the system and a few study on system robustness with \emph{inductive-based} (i.e., reasoning from observed training cases to general rules) transfer learning adaptations to data drift have been deployed \cite{Overview_AICSI} originally extend the practicality of CSI feedback systems.

\begin{figure}[htp]
    \centering
    \subfloat[Vanilla AE.]{
    \includegraphics[width=2.00 in]{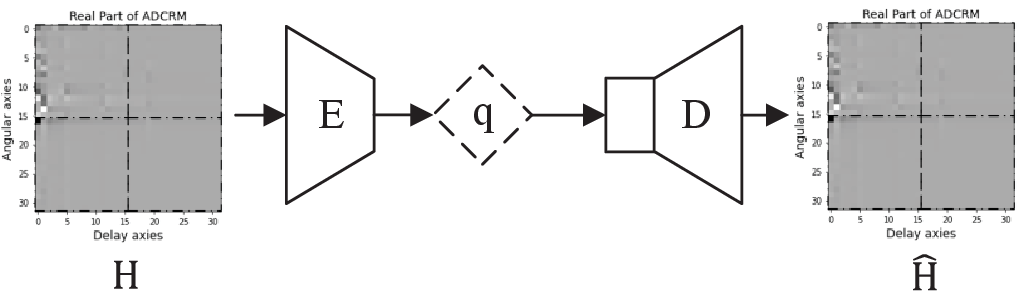}}
\quad
    \subfloat[HybridCVLNet Overview.]{
    \includegraphics[width=2.40 in]{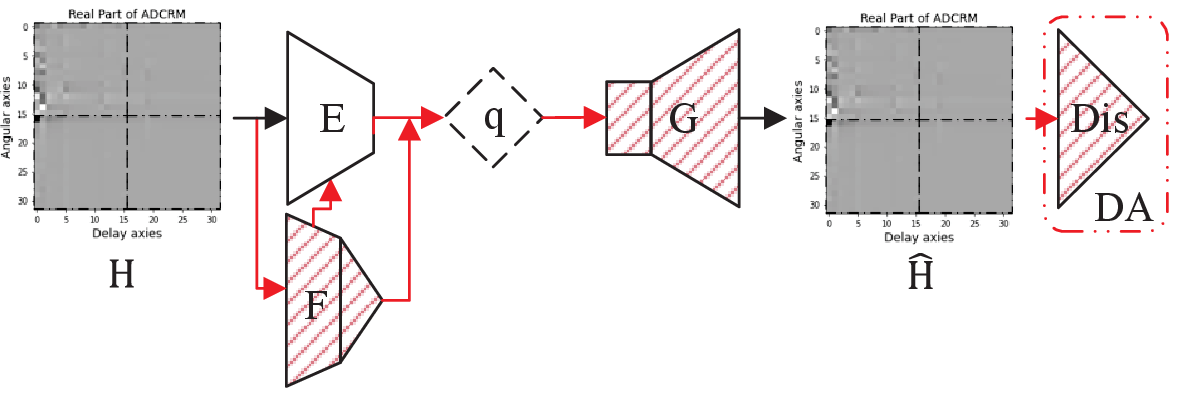}}
\caption{A comparison between the HybridCVLNet with the Vanilla  is demonstrated, where E and D stand for the encoder and decoder, respectively, and the wireless transmitting and quantization processes are summarized by q. The decomposed branch of the HybridCVLNet is noted as F. The generative decoder and discriminator in the domain adaptation stage are presented with G and Dis, respectively.}
\label{OV}
\end{figure}

In this paper, we propose a hybrid complex-valued lightweight CSI feedback system, namely the HybridCVLNet, based on the insight of neural network generalizability from the perspective of lightweight architecture and optimization, expressibility of the latent codeword and convolutional component, such that the impact of dataset bias in the intra-domain heterogeneous dataset can be mitigated. A comparison with vanilla AE architecture is illustrated in Fig. \ref{OV}. Subsequently, to address the challenge of data drift under inter-domain online optimization, a hybrid \emph{transductive-based} (i.e., reasoning from observed, specific training cases to specific test cases) domain adaptation framework is proposed in conjunction with the characteristics of HybridCVLNet.  

\begin{figure}[htp]
    \centering
    \subfloat[Feature DA.]{
    \includegraphics[width=0.40\linewidth]{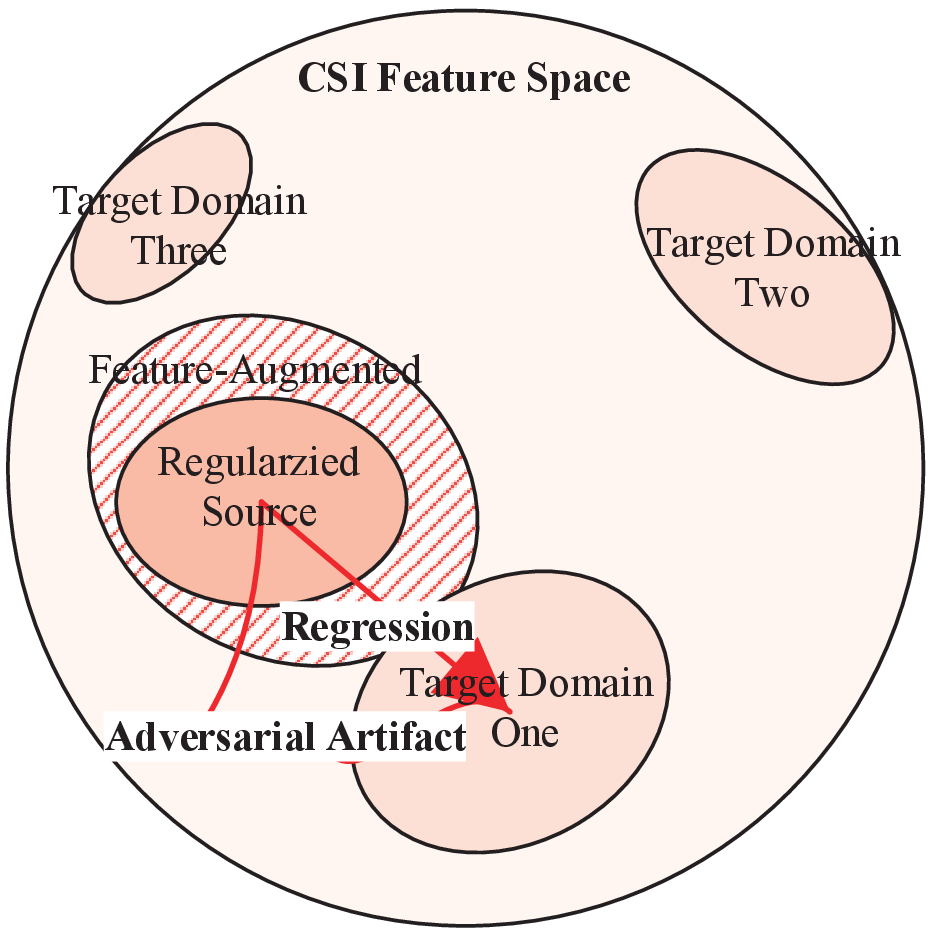}}
\quad
    \subfloat[Distribution DA.]{
    \includegraphics[width=0.40\linewidth]{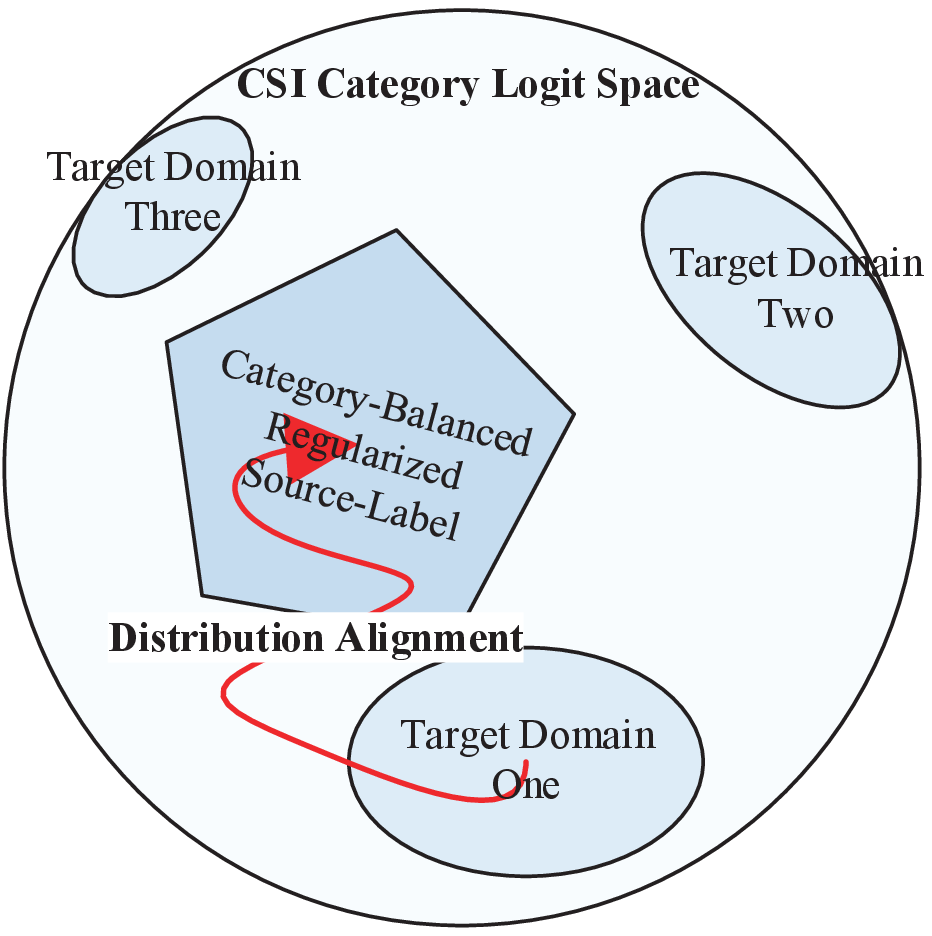}}
\caption{Illustration of feature space and category logits space domain adaptation, where the red slash marks augmented source domain in the feature space, and category-balancing source domain is represented by iso-polygon in the category logit space.}
\label{DA_OV}
\end{figure}

Specifically, this paper makes the following contributions:
\begin{itemize}
\item{\textbf{Hybrid Architecture $\&$ Task}: To overcome the dataset bias, the HybridCVLNet involves regularization during optimizing by decomposing the AE into two lightweight branches, appending a classifier to one of them, and end-to-end training with the combined objective of CSI regression and classification, as shown in of Fig. \ref{OV}.(b). Thus, the CSI reconstruction would be regularized by the discriminative task and prone not to be overfitted. Moreover, the discriminative branch features are believed to be category-related and complement another branch. Meanwhile, a category-balanced multi-label CSI dataset, as the regularized dataset, is proposed utilizing the relatively accessible standardized channel model.}
\item{\textbf{Hybrid Deterministic $\&$ Stochastic Codeword}: Intuitively, the HybridCVLNet introduces regularized classification that conflicts with the primary regression task in joint optimization. Thus, we incorporate the statistical categorical information into the reconstruction to compensate for the primary regression performance. We reorganize the reporting codeword into the sample-level channel scatterer indication, the deterministic category-related and complementary features, and the stochastic classification logits. Meanwhile, we integrate the self-attention mechanism with the complex-valued convolution in an interpretable manner.}
\item{\textbf{Hybrid Feature $\&$ Distribution Domain Adaptation}: Tackling the data drift to the target domain during inter-domain online fine-tuning, a hybrid transductive-based domain adaptation framework is proposed based on the HybridCVLNet with the proposed regularized dataset as the source domain. In the CSI feature space, an adversarial generative stimulation is adopted to facilitate pattern regression from the source to the target domain, illustrated in Fig. \ref{DA_OV}(a). In the CSI category logits space, a multi-label category distribution alignment is embraced to constrain the validity of category-related functions in the target domain utilizing the prior statistics of the category-balanced source domain, shown in Fig. \ref{DA_OV}(b).}
\end{itemize}

\section{Related Work}
In this section, literature that related to our contributions from the perspective of CSI feedback regularization, the reporting components and the module representativity, inductive-based transfer learning and transductive-based domain adaptation framework are enumerated. 
 
Regularization is an effective solution to dataset bias. In disparity with the vanilla AE, the distribution estimation framework, for instance,  generative adversary networks (GANs) \cite{GAN_CSI} and variational autoencoder (VAE) \cite{VAE_CSI} both deployed in the stochastic fertile space to reconstruct CSI with the adversarial objective under the distribution divergence regularization. Theoretically, GAN and VAE with stochastic codewords are less sensitive to dataset bias. However, these schemes fetch extensive model parameters and higher computational complexity as the challenge of regressing from distribution to image-level features. Another way to regularize the regression is to introduce related learning tasks. For instance, the classification has been preliminarily introduced into CSI feedback as a channel environment indicator \cite{Sensing} and an optimal compression rate selector \cite{AdNet} as a pre-training module. These methods isolatedly map features to manually defined labels, refine the dataset into homogeneous subsets and train proprietary sub-systems with class consistency to obtain performance gain. Nevertheless, they ignore the categorical regularization with classification. By contrast, our proposed HybridCVLNet draws on the insensitivity of the statistical latent to dataset bias and jointly optimizes conflicting tasks through a hybrid structure and learning objective.

From the view of CSI reporting composition and module representativity, plenty of studies transmit critical features about the downlink channel, such as the computed precoding matrices that are inspired by CSI reporting conducted in engineering \cite{Overview_AICSI}. However, the ambiguous side-information to feature transformation raises computational overhead and may disturb the full-precision downlink channel matrices regression objective. So far, extensive works have exploited the representativity of network depth \cite{CSINet_Plus}, width \cite{MRFNet}, and cardinality \cite{CRNet} of the CSI feedback system. The fitness of the complex-valued convolutional to the CSI was consolidated \cite{CV_3DCNN}. Nevertheless, the performance-superior self-attention mechanisms have yet to be merged with complex-valued convolution. Thus, we reorganize the reporting codeword with a low-complexity channel scatterer indication and semantic category latent. An interpretable complex-valued self-attention mechanism that is equivalent to the delayed tap magnitude value of the wireless channel is derived.

 A growing body of work has proposed many evolutions to inductive-based transform learning frameworks that leverage online \cite{online}, meta or multi-task \cite{Multi-task}, and federal learning \cite{Overview_AICSI} that directly reduce the risk of performance crashing in the data drift. However, existing inductive-based transfer optimization disregards the optimization direction from the prior (source domain) to the posterior distribution (target domain). The transductive-based domain adaptation scheme is a well-established approach to solving data drift in the CV field. Existing work maps the distribution-different source and target domains from a common latent representation in generative space to optimize \cite{GDA_2021_CVPR}. However, to our knowledge, the domain adaptation scheme has yet to be utilized over the CSI feedback. To fully utilize HybridCVLNet in terms of structure and learning objectives, the proposed hybrid domain adaptation scheme combined with feature-space representation and category logits-space distribution can achieve better performance on the target domain and be more efficient.

\section{System Model}

In FDD massive MIMO system, we consider a single-cell scenario where the BS equipped with a uniform linear antenna array (ULA) with $N_t$ transmit antennas among $N_c$ subcarriers, which serves a single-antenna UE ($N_r$ = 1). In the downlink transmission, the received signal at the UE is given as
\begin{equation}
\label{Downlink_reciver}
\mathbf{y}_{\text{multi}}=\mathbf{H}_{\text{multi}}\mathbf{s}+\mathbf{v},
\end{equation}
where $\mathbf{y}_{\text{multi}} = \left[y_1,y_2,\ldots,y_{N_{t}}\right]^T\in\mathbb{C}^{N_{t}\times1}$, $\mathbf{s}\in\mathbb{C}^{N_c\times1}$ indicates the transmitted pilot signal of the BS, and $\mathbf{v}=[v_1,v_2,\ldots,v_{N_{t}}]^T\in\mathbb{C}^{N_{t}\times1}$ is the vector of complex-valued additive white Gaussian noise, each element of which is independent and identically distributed (i.i.d) following $\mathcal{CN}\left(0,\sigma^2\right)$, $\mathbf{H}_{\text{multi}}=\left[\mathbf{h}_1,\mathbf{h}_2,\ldots,\mathbf{h}_{N_{t}}\right]^H\in\mathbb{C}^{{N_c\times N}_t}$ is the complex-valued downlink channel matrix.

The angular-delay channel response matrix (ADCRM) $\mathbf{H}_{\text{AD}}$ has a sparse property comes from a 2-dimensional (2-D) normalized discrete Fourier transform (DFT) in ULA, denoted as ${\mathbf{H}_{\text{AD}}}=\mathbf{F}_{\text{a}}\mathbf{H}_{\text{multi}}\mathbf{F}_{\text{d}}^H=[h_{\theta_1},h_{\theta_2},\ldots,h_{\theta_N}]^H\in\mathbb{C}^{\tau_N\times\theta_N}$, where $\mathbf{F}_{\text{a}}\in\mathbb{R}^{N_t\times\theta_N}$ is the normalized DFT matrix for space-to-angular domain transformation, $\theta_N$ is the angular domain resolution and each element of 
$\mathbf{F}_{\text{a}}$ is defined as ${\mathbf{F}_{\text{a}}\left[k,n\right]}_{k\in\left(1,...,N_t\right),n\in\left(1,...,\theta_N\right)}=\ {\frac{1}{\sqrt{N_t}}e}^{-j2\pi kn/N_t}$. Similarly, the $\mathbf{F}_{\text{d}}\in\mathbb{R}^{N_c\times\tau_N}$ denotes the normalized DFT matrix from frequency-to-delay domain transformation, and $\tau_N$ is the delay sample points. Due to the limited propagation delay of the path, only the first $\tau^\prime_N$ rows of ${\mathbf{H}_{\text{AD}}}$ contain non-zero information, thus ${\mathbf{H}_{\text{AD}}}$ can be truncated to ${\mathbf{H}^\prime_{\text{AD}}}$ losslessly, where ${\mathbf{H}^\prime_{\text{AD}}}\in\mathbb{C}^{\tau_N^\prime\times\theta_N}$. Each element of ADCRM ${\mathbf{H}^\prime_{\text{AD}}}$ corresponds to a certain path delay $\tau_{n_p}$ of path $n_p$, with a certain angle of arrival (AOA) indicated as $\theta_{n_p}$ can be written as
\begin{equation}
\label{delay_response}
\mathbf{h}_{\text{down}}(t)=\sum_{n_p=1}^{N_p}{f_{n_p}(t)\mathbf{a}_{\text{down}}(\theta_{n_p})d(t-\tau_{n_p})},
\end{equation}
where $N_p$ is the number of physical paths, $f_{n_p}(t)$ is the time-varying fading coefficient associated with the $n_p$-th path at moment $t$. Furthermore, $\mathbf{a}_{\text{down}}(\theta_{n_p})\in\mathbb{C}^{N_t\times1}$ represents the downlink steering vector of $n_p$-th path. In a complex-valued neural network, the complex-valued ADCRM is represented by a single channel complex-valued tensor $\mathfrak{R}(\mathbf{h}_{\text{down}}(t))+\mathfrak{I}(\mathbf{h}_{\text{down}}(t))\cdot{j}$, where $\mathfrak{R}(\mathbf{h}_{\text{down}}(t))$ and $\mathfrak{I}(\mathbf{h}_{\text{down}}(t))$ are the real and imaginary part of ADCRM, respectively.

Let $\mathbf{s}$ denote the $S$-dimensional vector of the compressed codeword, which is the output of the UE side encoder with mapping $E$. We can formulate the compression as follows:
\begin{equation}
\label{EN_Mapping}
\mathbf{s} = E(\mathbf{H}^\prime_{\text{AD}},W_\gamma^{E}),
\end{equation}
where $W_\gamma^{E}$ is the weight of the specific compression rate (CR) $\gamma$ of the deep compression setting, the compression rate $\gamma$ is defined as $\gamma = \frac{S}{2\times\tau_N^\prime\times{N_{t}}}$. Once the BS receives the codeword $\mathbf{s}$, the decoder with transformation $D$ will perform the downlink channel reconstruction by 
\begin{equation}
\label{DE_Mapping}
\widehat{\mathbf{H}}^\prime_{\text{AD}}= D(\mathbf{s},W_\gamma^{D}).
\end{equation}

In the typical domain generalization (DG) setting, there are $N$ multiple source domains, ${\mathbb{D}^1,...,\mathbb{D}^N}$, where each $\mathbb{D}^j$ with joint distribution ${\mathbb{P}^j(\mathcal{X},\mathcal{Y}):\left\{x\in\mathcal{X}, y\in\mathcal{Y}\right\}}_{j=1}^M$ contains $M$ data and label pairs. DG aims to infer on target domain $\mathbb{D}^{N+1}$ that $\mathbb{P}^j(\mathcal{X},\mathcal{Y})\neq \mathbb{P}^{N+1}(\mathcal{X},\mathcal{Y}), j=1, ..., N$. Considering a training model of CSI feedback with a feature extractor $E_\theta$ and a decoder $D_\psi$, parameterized by $\theta$ and $\psi$, respectively. Empirical Risk Minimization (ERM) is a definition that can perform as an objective measurement of the generalizability and stability of the dataset bias of the statistical learning or neural network \cite{SFA}. The model $D_\psi\circ E_\theta$ fitting the ERM can be defined as follow:
\begin{multline}
\label{ERM_system}
\text{arg}\underset{\theta,\psi}{\text{min}}\mathcal{L}_\text{erm}(\{\mathbb{D}^j\},E_\theta, D_\psi)\Longrightarrow\\
\text{arg}\underset{\theta,\psi}{\text{min}}\frac{1}{N}\sum_{j=1}^{N}\frac{1}{|\mathbb{D}^j|}\sum_{\mathbf{H}^\prime_{\text{AD}},\widehat{\mathbf{H}}^\prime_{\text{AD}}\in\mathbb{D}^j}\ell(E_\psi\circ D_\theta(\mathbf{H}^\prime_{\text{AD}}), \widehat{\mathbf{H}}^\prime_{\text{AD}}).
\end{multline}
$\ell$ is the corresponding loss function in the CSI feedback system. 

\section{The HybridCVLNet}
\begin{figure*}[!t]
    \centering
    \includegraphics[width=6.30 in]{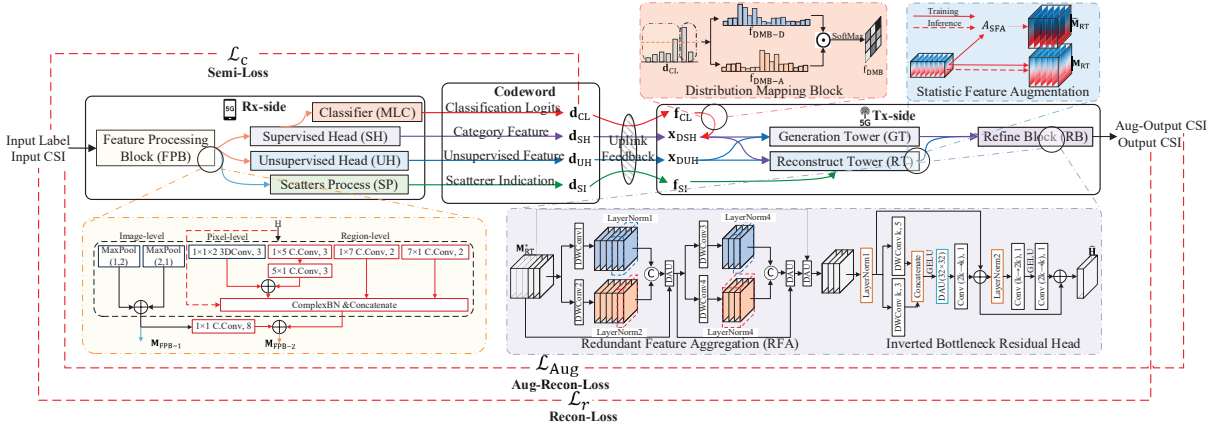}
     \caption{Overview of the HybridCVLNet.}
    \label{Over_OFFLine}
    \setlength{\belowcaptionskip}{-5pt}
\end{figure*}

In this section, We present the structure of HybridCVLNet and its learning objectives illustrated in Fig. \ref{Over_OFFLine}. The complex-valued Self-attention Mechanism is detailed, and insights into the regularized optimization are summarized. Defining a domain $\mathbb{D}_{XY}$ on input space $\mathcal{X}$ and label space $\mathcal{Y}$ with distribution by $\mathbb{P}$, the dataset $\mathcal{D}=\left\{\left(\mathbf{H}_i,\mathbf{y}_i\right)\right\}_{i=1}^{N}\sim\mathbb{P}$ samples from the domain, in which $N$ is the dataset size. The objective of the HybridCVLNet is to find the optimal classification function $f_c:\mathcal{X}\rightarrow\mathcal{Y}$ and regression mapping $f_r:\widehat{\mathcal{X}}\rightarrow\mathcal{X}$ on $\mathcal{D}$ simultaneously, where $\widehat{\mathcal{X}}$ is the reconstructed feature from the HybridCVLNet.

\subsection{HybridCVLNet Model and Learning Objective} 
\subsubsection{HybridCVLNet Encoder (UE-side)}
To overcome the dataset bias of the intra-domain heterogeneous dataset, the proposed HybridCVLNet encoder primarily outputs latent $\mathbf{s}$ with regression loss (i.e., to achieve $f_r$) and, in the second place, to exports category results with the constraint of regularized classification loss (i.e., to achieve $f_c$) and makes up a CSI reporting with a hybrid composition.

\begin{figure}[hbtp]
    \centering
    \includegraphics[width=2.35 in]{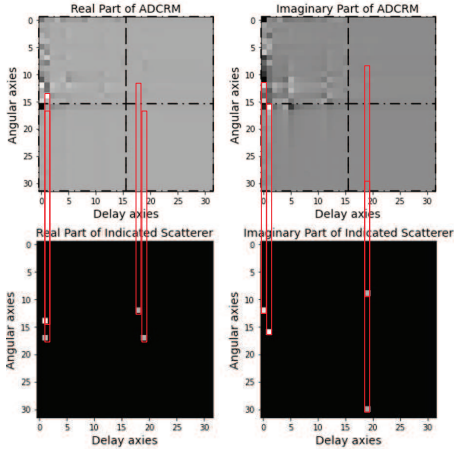}
     \caption{Illustration of Channel scatterer indication with peak response values and their locations. The red box indicates the $\text{argmax}(i_{\text{max},\bm{q}},j_{\text{max},\bm{q}})$ and the corresponding elements of $\mathbf{H}$.}
    \label{SI sample}
    \setlength{\belowcaptionskip}{-5pt}
\end{figure}

Given a complex-valued CSI $\mathbf{H}\in\mathbb{C}^{1\times n_t\times n_c}$, the Feature Enhance Block (FPB) of parameters $W_\text{FPB}$ generates informative representations $\mathbf{M}_{\text{FPB-2}}\in\mathbb{R}^{K_\text{FPB}\times n_t\times n_c}$ for further extraction, and the $\mathbf{M}_\text{FPB-1}\in\mathbb{C}^{2\times n_t\times n_c}$ emphasize the channel scatterers. To preserve sample-wise information in the compressed codeword \cite{DeepCompress1}, the HybridCVLNet introduces a low-complexity scatterer processing (SP) block that indicates and recalibrates scatterer along the downlink environment with output $\mathbf{d}_\text{SI}\in\mathbb{R}^{2\times8}$, which contains the peak response value of $\underset {i,j \in{\bm{q}_{(1,...,4)}}} {\text{max}}\mathbf{M}_{\text{FPB-1}_{i,j}}\in\mathbb{R}^{1\times8}$ and the peak value sequence $\text{argmax}(i_{\text{max},\bm{q}},j_{\text{max},\bm{q}})\in\mathbb{R}^{1\times8}$, demonstrated in Fig. \ref{SI sample}. The $\mathbf{d}_\text{SI}$ manually divides four quadrants $\bm{q}$ of the ADCRM by its far or near delay and small or large angle properties.

To achieve $f_c$, the top branch Supervised Header (SH) with function $g_{\text{SH}}$ of $W_{\text{SH}}$ is docked with a Multi-label Classifier (MLC) $g_{\text{MLC}}$ of parameters $W_{\text{MLC}}$ that produces CSI classification logits $\mathbf{d}_\text{CL}\in\mathbb{R}^7$ that can be formulated as,
\begin{equation}
\label{MLC_Mapping}
\mathbf{d}_\text{CL} = g_{\text{MLC}}(W_{\text{MLC}},g_{\text{SH}}(W_{\text{SH}},\mathbf{M}_\text{FPB-2})).
\end{equation}

Under the constraints of the classification objective, SH has category-related and class-specific output features $\mathbf{x}_{\text{SH}}$. To complement the discarded category-invariant details, the bottom branch Unsupervised Header (UH), with parameters $W_{\text{UH}}$ of mapping $g_{\text{UH}}$ generates the complementary features $\mathbf{x}_{\text{UH}}$ \cite{Hybrid}. Compressed deterministic codewords $\mathbf{d}_{\text{SH}}$ and $\mathbf{d}_{\text{UH}}$ are obtained with two real-valued dense layers $g_{\text{SH}_{\text{D}_\gamma}}$ and $g_{\text{UH}_{\text{D}_\gamma}}$ in compression rate setting $\gamma$ as follow,
\begin{equation}
\begin{aligned}
\label{SHUH}
\mathbf{d}_\text{SH} = g_{\text{SH}_{\text{D}_\gamma}}(\mathbf{x}_{\text{SH}})&, \;\mathbf{x}_{\text{SH}} = g_{\text{SH}}(W_{\text{SH}},\mathbf{M}_\text{FPB-2}),\\
\mathbf{d}_\text{UH} = g_{\text{UH}_{\text{D}_\gamma}}(\mathbf{x}_{\text{UH}}+\mathbf{x}_{\text{SH}})&,\;\mathbf{x}_{\text{UH}} = g_{\text{UH}}(W_{\text{UH}},\mathbf{M}_\text{FPB-2}),\\
\end{aligned}
\end{equation}
note that we aggregate $\mathbf{x}_{\text{SH}}$ and $\mathbf{x}_{\text{UH}}$ at UH to enrich features. Therefore, the downlink codeword reporting $\mathbf{s}$ with the length of $S$ has four components: the scatterer indication $\mathbf{d}_\text{SI}\in\mathbb{R}^{2\times8}$, the stochastic classification logits $\mathbf{d}_\text{CL}\in\mathbb{R}^{7}$, the deterministic category-related codeword $\mathbf{d}_\text{SH}\in\mathbb{R}^{(S/2)-7}$ and the complementary unsupervised codeword $\mathbf{d}_\text{UH}\in\mathbb{R}^{(S/2)-16}$, which can be formulated as
\begin{equation}
\label{Codeword}
\mathbf{s}=\{\mathbf{d}_\text{SI},\mathbf{d}_\text{CL},\mathbf{d}_\text{SH},\mathbf{d}_\text{UH}\}\in{\mathbb{R}^{S}}.
\end{equation}

\subsubsection{HybridCVLNet Decoder (BS-side)}
The proposed HybridCVLNet decoder primarily decompresses and restores the deterministic codewords $\mathbf{d}_\text{SH}$ and $\mathbf{d}_\text{UH}$ with calibration of indication $\mathbf{d}_\text{SI}$ (i.e., to achieve $f_r$). Meanwhile, to utilize the stochastical $\mathbf{d}_\text{CL}$, the decoder may implicitly map the reporting to regression-oriented features in
$f_s:\mathcal{Y}\rightarrow\mathcal{X}$ on dataset $\mathbb{D}$. 

The functionally differentiated processing of the hybrid reporting $\mathbf{s}$ is formulated as follows,
\begin{equation}
\begin{aligned}
\label{Codeword_Processing}
\mathbf{x}_{\text{DSH}}&=g_{\text{DSH}_{\text{D}_\gamma}}(\mathbf{d}_\text{SH}),\;  \mathbf{x}_{\text{DUH}}=g_{\text{DUH}_{\text{D}_\gamma}}(\mathbf{d}_\text{UH}),\\
\mathbf{f}_{\text{SI}}&=\text{MaxUnpooling}(\mathbf{d}_{\text{SI}}),\\
\mathbf{f}_{\text{CL}}&=\text{SoftMax}(g_{\text{CMI}}(W_{\text{CMI}},\mathbf{d}_{\text{CL}})),\\
\end{aligned}
\end{equation}
where two corresponding real-value dense layers $g_{\text{DSH}_{\text{D}_\gamma}}$ and $g_{\text{DUH}_{\text{D}_\gamma}}$ are proposed to preliminarily decompress the deterministic codewords. An indication $\mathbf{f}_{\text{SI}}\in\mathbb{R}^{\tau_N^\prime\times\theta_N}$ is transformed with the scatterer mapping indicator (SMI) with MaxUnpooling operation to recalibrate features. A category embedding score $\mathbf{f}_{\text{CL}}$ is obtained with a category mapping indicator (CMI) $g_{\text{CMI}}$ with parameter $W_{\text{CMI}}$. 

Theoretically, the transmitted $\mathbf{d}_{\text{CL}}$ can be regarded as a piece of high-dimensional semantic \cite{Hyper-code}. HybridCVLNet decoder introduces a two-branch delay and angular Distribution Mapping Block (DMB). The $g_{\text{DMB}}$ with $W_{\text{DMB-D}}$ and $W_{\text{DMB-A}}$ that map the $\mathbf{d}_{\text{CL}}\in\mathbb{R}^7$ to a delay-direction and angular-direction distribution-related scores $\mathbf{f}_{\text{DMB-D}}\in\mathbb{R}^{\tau_{N}^\prime}$ and $\mathbf{f}_{\text{DMB-A}}\in\mathbb{R}^{\theta_N}$. A final correlation score $\mathbf{f}_{\text{DMB}}\in\mathbb{R}^{\tau_N^\prime\times\theta_N}$ can be defined as
\begin{multline}
\label{Distribution_Mapping}
\mathbf{f}_{\text{DMB}}=\mathbf{f}_{\text{DMB-D}}\cdot\mathbf{{f}_{\text{DMB-A}}}^T=\\\text{SoftMax}(g_{\text{DMB}}(\{W_{\text{DMB-D}},W_{\text{DMB-A}}\},\mathbf{d}_{\text{CL}})).
\end{multline}

The upper Generator Tower (GT) branch of parameters $W_{\text{GT}}$ provides more parameter space \cite{FishNet} in decompressing and refinement mapping $g_{\text{GT}}$. Another Reconstruct Tower (RT) branch $g_{\text{RT}}$ with simpler parameter space $W_{\text{RT}}$ rebuild the coarser character of the CSI. The output feature $\mathbf{M}_\text{GT}$ and $\mathbf{M}_\text{RT}$ are formulated as:
\begin{equation}
\begin{aligned}
\label{GTRT}
\mathbf{M}_\text{GT}&=g_{\text{GT}}(W_{\text{GT}},\text{Concat}\{\mathbf{x}_{\text{DUH}},\mathbf{x}_{\text{DSH}}\cdot\mathbf{f}_{\text{CL}}\},\mathbf{f}_{\text{DMB}}),\\
\mathbf{M}_\text{RT}&=g_{\text{RT}}(W_{\text{RT}},\text{Concat}\{\mathbf{x}_{\text{DUH}},\mathbf{x}_{\text{DSH}}\cdot\mathbf{f}_{\text{CL}}\})+\mathbf{M}_\text{GT}.\\
\end{aligned}
\end{equation}

Data augmentation can improve domain generalizability by creating variability of features and flexible model \cite{Tele_AUG}. However, image augmentation may disturb the directionality and destroy the category invariable of CSI \cite{Randaugment}. Moreover, it would crop, shuffle, and block the essential semantics of ADCRM. Thus, the HybridCVLNet decoder embeds a Statistic Feature Augmentation (SFA) block \cite{SFA} to RT that empirically improves the generalizability without destroying the image level feature of the CSI. The augmented SFA output $\widetilde{\mathbf{M}}_{\text{SFA}}\in\mathbb{C}^{K_\text{RT}\times\tau_N^\prime\times\theta_N}$ with augmentation function $A_{\text{SFA}}(\cdot)$ defines as 
\begin{equation}
\begin{aligned}
\label{SFA}
&\widetilde{\mathbf{M}}_{\text{SFA}}=A_{\text{SFA}}(\mathbf{M}_{\text{RT}})=\mathbf{\alpha}\odot \mathbf{M}_{\text{RT}}+\mathbf{\beta},\\
&\mathbf{\alpha}\sim\mathcal{N}(1,\sigma_\alpha\mathbf{I}), \mathbf{\beta}\sim\mathcal{N}(0,\sigma_\beta\mathbf{I}),\\
\end{aligned}
\end{equation}
where $\mathbf{\alpha}, \mathbf{\beta}\in\mathbb{R}^{K_\text{RT}\times n_t\times n_c}$ are the noise components sampling from two multivariate Gaussian distribution $\mathcal{N}(\mu,\Sigma)$ in this paper. Empirically, we set the statistical moment $\mu_\alpha=1$ and $\mu_\beta=0$, $\Sigma_\alpha=\sigma_\alpha\mathbf{I}$ and $\Sigma_\beta=\sigma_\beta\mathbf{I}$ respectively, where $\mathbf{I}$ is the identity matrix, $\sigma_1, \sigma_2$ are two equal scalars.

Inspired by the SOTA image restoration method \cite{Light_RES}, a lightweight real-valued Refine Block (RB) \cite{Conv2020s} is appended to the HybridCVLNet decoder to boost and stabilize the reconstruct performance, as shown in the bottom left zoom-in block of Fig. \ref{Over_OFFLine}. Thus, for the real-valued $\mathbf{M}_{\text{RT}}^\ast,\;\widetilde{\mathbf{M}}_{\text{SFA}}^\ast\in\mathbb{R}^{{2K}_{\text{RT}}\times\tau_N^\prime\times\theta_N}$, the RB block in refinement function $g_{\text{RB}}$ with $W_{\text{RB}}$ obtains prediction $\widehat{\mathbf{H}}$ with its augmentation $\widetilde{\mathbf{H}}$ simultaneously, formulated as,
\begin{equation}
\label{RB_Mapping}
\widehat{\mathbf{H}}=g_{\text{RB}}(W_{\text{RB}},\mathbf{M}_{\text{RT}}^\ast+\mathbf{f}_{\text{SI}}),\widetilde{\mathbf{H}}=g_{\text{RB}}(W_{\text{RB}},\widetilde{\mathbf{M}}_{\text{SFA}}^\ast+\mathbf{f}_{\text{SI}}) .
\end{equation} 

\subsubsection{HybridCVLNet Learning Objective}
Suppose the inputs and label lie in $\mathbf{H}\subset\mathcal{X}$ and $\mathbf{y}\subset\mathcal{Y}$. Two types of error calculation $\ell_\text{c}:\mathcal{Y}\times\mathcal{Y}\rightarrow\mathbb{R}$ and $\ell_\text{r}:\mathcal{X}\times\mathcal{X}\rightarrow\mathbb{R}$ are proposed corresponding to the hybrid classification and reconstruction tasks $f_c$ and $f_r$ respectively, each objective component of HybridCVLNet is listed as follows.
\begin{equation}
\begin{aligned}
\label{Loss_One}
\mathcal{L}_\text{c}^N(W_{\text{C}})=&\sum_{i=1}^{N}\ell_\text{c}({\mathbf{d}_\text{CL}}_i,\mathbf{y}_i),\\
\mathcal{L}_\text{r}^N(W_{\text{R}})=\sum_{i=1}^{N}\ell_\text{r}(\widehat{\mathbf{H}}_i,\mathbf{H}_i),\;&
\mathcal{L}_{\text{Aug}}^N(W_{\text{R}})=\sum_{i=1}^{N}\ell_\text{r}(\widetilde{\mathbf{H}}_i,\mathbf{H}_i),\\
\end{aligned}
\end{equation}
where $W_{\text{C}}=\{W_{\text{FPB}},W_{\text{SH}},W_{\text{MLC}}\}$ is set of weights participate into the task $f_c$, $W_{\text{R}}=\{W_{\text{FPB}},W_{\text{SH}},W_{\text{UH}},W_{\text{GT}},W_{\text{RT}},W_{\text{RB}}\}$ is set of whole HybridCVLNet parameters in task $f_r$. It is instinctive to achieve joint optimization by summing up these objectives. However, a critical challenge of HybridCVLNet is the reconstruction term $\mathcal{L}_\text{r}$ arguably conflicts with the classification term $\mathcal{L}_\text{c}$ \cite{Hybrid}. During trials, we observed that $\mathcal{L}_\text{c}$ converges faster than $\mathcal{L}_\text{r}$. During trials, we observed that $\mathcal{L}_\text{c}$ converges faster than $\mathcal{L}_\text{r}$. Meanwhile, $\mathcal{L}_\text{c}$ in smooth convergence is higher than $\mathcal{L}_\text{c}$ due to the different error calculations. We have also tried introducing SOTA multi-task learning mechanisms. fot instance, the gradient normalization \cite{GardN} and uncertainty modelling \cite{Multi_Uncertain}. 

However, these approaches are inefficient compared to hyperparameter-weighted loss functions since the output of $\mathcal{L}_\text{c}$ and $\mathcal{L}_\text{r}$ are different in terms of the depth. Meanwhile, the difference between the two objective values is relatively fixed. Thus, we define the total objective function of the HybridCVLNet as
\begin{equation}
\label{HybridCVLNet_Objective}
\mathcal{L}_{\text{Offline}}^N=[\mathcal{L}_{\text{r}}^N+\mathcal{L}_{\text{Aug}}^N\cdot\delta(t-\bm{Epoch}_{\text{Aug}})]+\lambda_\text{Rug}\cdot\mathcal{L}_{\text{c}}^N,
\end{equation}
where $\bm{Epoch}_{\text{Aug}}$ is defined as a set of epochs indicator, which accounts for the effectiveness of the $\mathcal{L}_{\text{Ar}}$. The $\lambda_\text{Rug}$ is defined to balance the $\mathcal{L}_{\text{r}}$ and $\mathcal{L}_{\text{c}}$ terms. The optimization strategy of HybridCVLNet is to promote simultaneous convergence of regression and classification tasks with $\mathcal{L}_{\text{r}}$ and $\mathcal{L}_{\text{c}}$ at the same magnitude by offsetting the higher term and add the constraint of $\mathcal{L}_{\text{Aug}}$ after $\mathcal{L}_{\text{r}}$ is stabilized.

\subsection{Embedding Complex-valued Self-attention Mechanism}
\begin{figure}[hbtp]
    \centering
    \includegraphics[width=3.00 in]{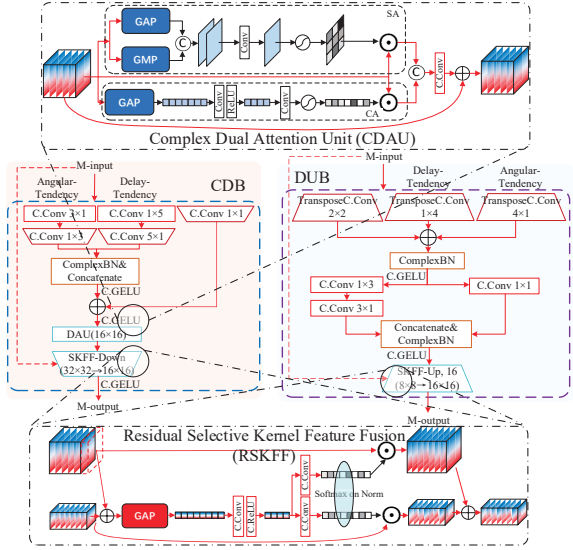}
     \caption{The Convolutional Down-sampling Block (CDB) and Deconvolutional Up-sampling Block (DUB) with Complex Dual Attention Unit (CDAU) Residual Selective Kernel Feature Fusion (RSKFF) utilizing the Complex-valued Self-attention mechanism.}
    \label{Complex_Micro}
    \setlength{\belowcaptionskip}{-5pt}
\end{figure}

The HybridCVLNet adopts the cascading complex-valued Convolutional Down-sampling Block (CDB) and Deconvolutional Up-sampling Block (DUB) that can efficiently embed phase information into the latent with the inductive bias of translation equivariance \cite{Conv2020s}, illustrated in Fig. \ref{Complex_Micro}. The self-attention mechanism weighs the input feature by defining how important (i.e. attention scores) an element is and from which dimensions (i.e. interactions) to judge. The HybridCVLNet derives an attention-score calculation for complex-valued features with interpretability and proposes multiple methods to apply it.

Given a complex-valued feature map $\mathbf{M}\in\mathbb{C}^{K\times H\times W}$, the self-attentiveness of element $\mathbf{m}_{h,w}$ of tensor $\mathbf{M}$ is quantified by the L2 norm value $\|\mathbf{m}_{h,w}\|_{2}$. The attentiveness can be utilized to calculate the attention scores, formulated as 
\begin{equation}
\label{Self-Attention_Weight}
\|\mathbf{m}_{h,w}\|_{2}=\sqrt[2]{{\mathfrak{R}(\mathbf{m}_{h,w})}^2+{\mathfrak{I}(\mathbf{m}_{h,w})}^2},
\end{equation}
that is regarded as a transformation from two discrete variables $\mathfrak{R}(\mathbf{m}_{h,w})$ and $\mathfrak{I}(\mathbf{m}_{h,w})$ to the distance away from the original point in the complex plane. Considering the sparsity of complex-valued ADCRM, the distance $\|\mathbf{m}_{h,w}\|_{2}$ intuitively describes the intensity of element $\mathbf{m}_{h,w}$. To increase the interpretability, the $\|\mathbf{m}_{h,w}\|_{2}$ is equivalent to the tap amplitude of wireless channel delay profile $|\mathbf{h}(t,\tau)|$, where $\mathbf{h}(t,\tau)=\sum_{i=1}^{N}{c_i\left(t\right)\delta\left(\tau-\tau_i\right)}$. Note that the impulse response $\mathbf{h}$ and the multipath component coefficient $c_i\left(t\right)$ varies with time $t$, representing the channel by a delay line with $N$ taps. 

The complex-valued self-attention on $\mathbf{M}$ can be exploited either with inter-channel or with inter-spatial interactions. A global pooling operation obtains a descriptor $\mathbf{d}_{\text{inter-channel}}\in\mathbb{R}^{K\times1\times1}$ or $\mathbf{d}_{\text{inter-spatial}}\in\mathbb{R}^{1\times H\times W}$ adopts on $\|\mathbf{m}_{h,w}\|_{2}$ across spatial dimension or channel dimension, which is lately activated through a convolutional layer followed by the Sigmoid gating. Motivated by the SOTA self-attention application \cite{CBAM}, the HybridCVLNet embeds a Complex Dual Attention Unit (CDAU), as shown in the upper zoom-in area of Fig. \ref{Complex_Micro} that suppresses less-useful while passing the more informative features. To promote multi-scale features interaction and receptive field variation of the  \cite{Rich_Res}, a Residual Selective Kernel Feature Fusion (RSKFF) block with multi-scale fusing and inter-channel self-attention is proposed, illustrated in the lower zoom-in area of Fig. \ref{Complex_Micro}.

\subsection{Standardized Regularization}
\subsubsection{Standardized Regularization Dataset}
\begin{table*}[!ht]
\captionsetup{size=footnotesize}
\caption{Delay and Angular Spread profile to enhance the CDL Data Diversity.} 
\label{CDL_DATA_Diversity}
\centering
\resizebox{0.75\linewidth}{!}{
\begin{tabular}{|llc|}
\hline
\multicolumn{3}{|c|}{Regularized CDL Parameters of Data Diversity}                                                                                                                            \\ \hline
\multicolumn{1}{|c|}{\multirow{3}{*}{Delay Scaling Factor (${\text{DS}}_{\text{desired}}$)}} & \multicolumn{1}{c|}{Short-delay profile} & 20ns(Indoor office),  45,65ns(UMi Street-canyon), 93ns(UMa) \\ \cline{2-3} 
\multicolumn{1}{|c|}{}                                        & \multicolumn{1}{c|}{Normal-delay profile} & 39ns(Indoor office), 129ns(UMi Street-canyon), 240ns(UMi/UMa O2I)  \\ \cline{2-3} 
\multicolumn{1}{|c|}{}                                        & \multicolumn{1}{c|}{Long-delay profile}   & 59ns(Indoor office), 316ns(UMi Street-canyon), 153ns(RMa\&RMa O2I) \\ \hline
\multicolumn{1}{|l|}{\multirow{4}{*}{Angular Scaling Factor (${\text{AS}}_{\text{desired}}$)}} & \multicolumn{1}{l|}{AOD spread ASD}       & 5.0\degree, 15.0\degree, 25.0\degree                                              \\ \cline{2-3} 
\multicolumn{1}{|l|}{}                                        & \multicolumn{1}{l|}{AOA spread ASA}       & 15.0\degree, 30.0\degree, 45.0\degree                                            \\ \cline{2-3} 
\multicolumn{1}{|l|}{}                                        & \multicolumn{1}{l|}{ZOA spread ZSA}       & 1.0\degree, 3.0\degree                                                    \\ \cline{2-3} 
\multicolumn{1}{|l|}{}                                        & \multicolumn{1}{l|}{ZOD spread ZSD}       & 1.0\degree, 3.0\degree, 5.0\degree, 10.0\degree                                          \\ \hline
\end{tabular}
}
\end{table*}

Vendors and Operators propose localized CSI feedback dataset In the latest 3rd Generation Partnership Project (3GPP) approval and discussion \cite{AICSI_in_Industry}. However, the CSI samples are so vast with uncountable image patterns that it is unrealizable to build a common dataset. Moreover, over-the-air collections and manual labelling are huge expenditures and consumption. Therefore, the regularization dataset is proposed based on two insights. One is that it should be easy to acquire or preferably standardized. Second, the dataset should be category-balanced and richly diverse for subsequent incremental learning.

\begin{figure}[ht]
    \centering
    \includegraphics[width=2.65 in]{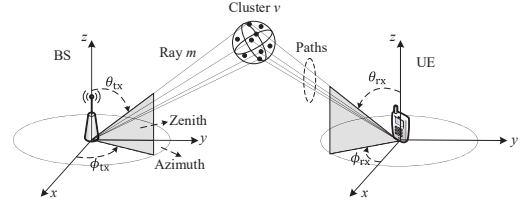}
     \caption{Clustered Delay Line Channel Modeling.}
    \label{CDL_MODEL}
    \setlength{\belowcaptionskip}{-5pt}
\end{figure}

A standardized 3-dimensional (3D) channel model was proposed in 3GPP TR 38.901, refers as the clustered delay line (CDL) channel model. Three CDL settings, namely CDL-A, B, and C, are constructed to represent channel profiles for non-line-of-sight (NLOS) environments, while CDL-D and E are for line-of-sight (LOS) environments. The CDL complex-valued channel is characterized by instantaneous multipath fading based on the UE moving speed, average signal power, delay time, random phase, and path azimuth and zenith angles of departure and arrival (AoD, ZoD, AoA and ZoA), which is illustrated in Fig. \ref{CDL_MODEL}. 
\vspace{-0.10cm}
\vspace{-0.08cm}

To enhance the diversity of the regularized dataset, scaled channel delay and angle spreads are involved in each CDL profile. The desired delay spread $\tau_{n,\text{scaled}}$ and ray angular profile $\phi_{n,\text{scaled}}$ can be scaled as
\begin{equation}
\begin{aligned}
\label{CDL_Diversity}
&\tau_{n,\text{scaled}}=\tau_{n,\text{model}}\cdot{\text{DS}}_{\text{desired}},\\
\phi_{n,\text{scaled}}=&\frac{{\text{AS}}_{\text{desried}}}{{\text{AS}}_{\text{model}}}(\phi_{n,\text{model}}-\mu_{\phi,\text{model}})+\mu_{\phi,\text{desired}},\\
\end{aligned}
\end{equation}
where normalized delay $\tau_{n,\text{model}}$ of the $n$-th cluster is scaled by delay spread ${\text{DS}}_{\text{desired}}$. The tabulated ray angle $\phi_{n,\text{model}}$ is the offset of the normalizaed angular ${\text{AS}}_{\text{model}}$. $\mu_{\phi,\text{model}}$, $\mu_{\phi,\text{desired}}$ are the mean and desired angles of tabulated CDL. We select 12 enumerated ${\text{DS}}_{\text{desired}}$ and angles in Table 7.7.3-2 of 3GPP TR 38.901, shown in Table \ref{CDL_DATA_Diversity}. A one-hot coding multi-labels classification (MLC) for the regularization dataset is established considering the delay profile (CDL-A/B/C/D/E) and propagation environment (LOS and NLOS), shown in Table \ref{labelcoding}.
\begin{table}[!htb]
\captionsetup{size=footnotesize}
\caption{Label Coding of CSI Classification.} \label{labelcoding}
\setlength\tabcolsep{0pt} 
\footnotesize\centering

\smallskip
\resizebox{0.45\linewidth}{!}{
\begin{tabular*}{0.55\columnwidth}{@{\extracolsep{\fill}}cccc}
\toprule
\multicolumn{2}{c}{CSI Label}                       & \multicolumn{2}{c}{\multirow{2}{*}{Class Coding}} \\ 
\multicolumn{1}{c}{Delay} & \multicolumn{1}{c}{Env} &  \\
\midrule
  CDL-A& NLOS& \multicolumn{2}{c}{[1, 0, 0, 0, 0, 0, 1]} \\
  CDL-B& NLOS& \multicolumn{2}{c}{[0, 1, 0, 0, 0, 0, 1]} \\
  CDL-C& NLOS& \multicolumn{2}{c}{[0, 0, 1, 0, 0, 0, 1]} \\
  CDL-D& LOS& \multicolumn{2}{c}{[0, 0, 0, 1, 0, 1, 0]} \\
  CDL-E& LOS& \multicolumn{2}{c}{[0, 0, 0, 0, 1, 1, 0]} \\
\bottomrule
\end{tabular*}
}
\end{table}

\subsubsection{Regularized Multi-label Classification}
\begin{figure}[hbtp]
    \centering
    \includegraphics[width=2.65 in]{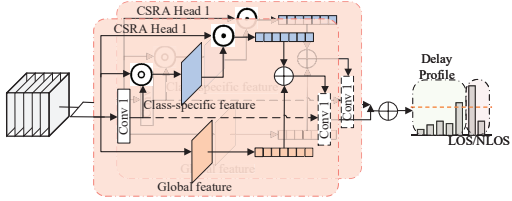}
     \caption{Class-Specific Residual Attention Module.}
    \label{CSRA_figure}
    \setlength{\belowcaptionskip}{-5pt}
\end{figure}

The HybridCVLNet adopts a class-specific residual attention (CSRA) module \cite{CSRA} to utilize the spatial self-attention for each object class, which can be viewed as a class-specific attention mechanism that focuses on category probability in the space of feature, shown in Fig. \ref{CSRA_figure}.
\vspace{-0.08cm}

Given a real-valued SH block output feature $\mathbf{M}_{\text{SH}}\in\mathbb{R}^{K_{\text{SH}}\times H_{\text{SH}}\times W_{\text{SH}}}$ that can be decoupled as $\mathbf{m}_j\in\mathbb{R}^{K_{\text{SH}}}$, where $j\in\mathbb{R}^{H_{\text{SH}}\times W_{\text{SH}}}$ means the $j$-th location of tensor $\mathbf{M}_{\text{SH}}$. A fully connected ($1\times1$ convolution) layer with $\mathbf{W}_{\text{MLC}_i}\in\mathbb{R}^{K_{\text{SH}}}$ is the classifier for the $i$-th class. We define the class-specific attention scores $\mathbf{u}_j^i$ for the $i$-th class and $j$-th location as
\begin{equation}
\label{CSRA-attention score}
\mathbf{u}_j^i=\frac{exp(T\mathbf{m}_j^T\odot\bm{W}_{\text{MLC}_i})}{\sum_{k=1}^{H_{\text{SH}}\times W_{\text{SH}}}{exp(T\mathbf{m}_k^T\odot\bm{W}_{\text{MLC}_i})}},
\end{equation}
where $\sum_{j=1}^{H_{\text{SH}}\times W_{\text{SH}}}\mathbf{u}_j^i=1$ and $T$ is the temperature hyperparameter controlling the sharpness of the score. We can view $\mathbf{u}_j^i$ as the probability of the class $i$ appearing at location $j$, corresponding to a data likelihood term hinge on both feature $\mathbf{M}_{\text{SH}}$ and classifier $\bm{W}_{\text{MLC}}$. The class-related feature score that enhanced to the feature $\mathbf{M}_{\text{SH}}$ is $\mathbf{w}\in\mathbb{R}^{1\times H_{\text{SH}}\times W_{\text{SH}}}$, where $\mathbf{w}=\sum_{i=1}^{k}\mathbf{u}_j^i$ is equivalent with the accumulate of class-related feature appearing probability at location $j$. 

The CSRA discriminate feature ${\mathbf{M}}_{\text{CSRA}}^i$ for the $i$-th class is a weighted combination of attention scores $\mathbf{a}_i$ for the $i$-th category on position $j$, and global class-agnostic feature vector $\mathbf{g}$, formulated as (\ref{CSRA})
\begin{equation}
\begin{aligned}
\label{CSRA}
\mathbf{a}^i=\sum_{j=1}^{H_{\text{SH}}\times W_{\text{SH}}}{\mathbf{u}_j^i\mathbf{m}_j},\: &\mathbf{g}=\frac{1}{H_{\text{SH}}\times W_{\text{SH}}}\sum_{j=1}^{H_{\text{SH}}\times W_{\text{SH}}}\mathbf{m}_j,\\
{\mathbf{M}}_{\text{CSRA}}^i&=\:\mathbf{g}+Lambda\ast\mathbf{a}^i.\\
\end{aligned}
\end{equation}

Finally, class-specific feature vectors are sent to the classifier to obtain the final logits $\mathbf{d}_{\text{CL}}\in\mathbb{R}^k$, and we further propose a simple multi-head attention extension to CSRA to tune the temperatures $T$. Multiple residual attention branches (or heads) are used, each utilizing a different temperature $T$. We denote the number of heads as $H$. The logits from different heads are added to get the final logits $\mathbf{d}_{\text{CL}}$, as
\begin{equation}
\begin{aligned}
\label{MLC_out}
\mathbf{y}_{T_i}\triangleq(\mathbf{y}^1,\mathbf{y}^2,...,\mathbf{y}^i)=(\mathbf{W}_{\text{MLC}_1}^T&{\mathbf{M}}_{\text{CSRA}}^1,...,\mathbf{W}_{\text{MLC}_i}^T{\mathbf{M}}_{\text{CSRA}}^i),\\
\mathbf{d}_{\text{CL}}=&\sum_{h=1}^{H}{\mathbf{y}}_{T_i}.\\
\end{aligned}
\end{equation}

\begin{figure*}[htbp]
    \centering
    \includegraphics[width=6.50 in]{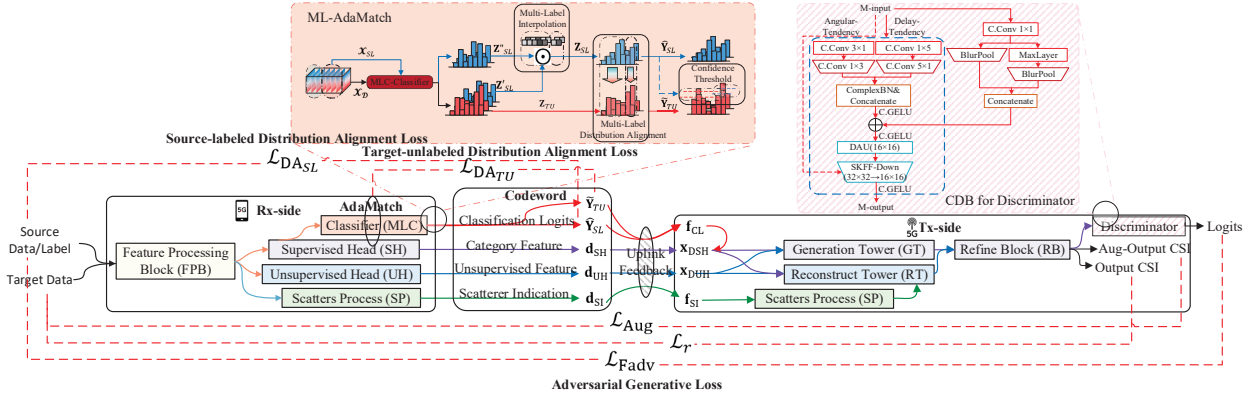}
     \caption{Overview of the HybridCVLNet Domain Adaptation.}
    \label{Overview_DA}
    \setlength{\belowcaptionskip}{-5pt}
\end{figure*}

\section{Domain Adaptation with HybridCVLNet}
This section details the proposed transductive-based category logits-space distribution alignment and the feature-space adversarial generative domain adaptation (DA) schemes. The hybrid domain adaptation (HDA) framework and its optimization strategy are demonstrated. The hybrid DA framework is shown in Fig. \ref{Overview_DA}.

\subsection{CSI Category Logits-space Distribution Alignment}
When inter-domain data drift appears, preserving the validity of the HybridCVLNet discriminative branch, i.e., the prior regularity information, can be helpful. The AdaMatch \cite{AdaMatch} is a categorical DA scheme that constrains the source and target domain logits $\mathbf{d}_{\text{CL},SL}$ and $\mathbf{d}_{\text{CL},TU}$ with distribution alignment, which encourages the target pseudo-labels in target distribution $\mathbb{P}_{TU}$ to follow the source label distribution $\mathbb{P}_{SL}$. We conduct an MLC version of AdaMatch (ML-AdaMatch) so that the target SH features $\mathbf{x}_{\text{SH},SL}$ are still category-related, and $\mathbf{d}_{\text{CL},TU}$ is valid, while the target domain is unknown.

We concatenate the source and target domain batch together, denoted as $\mathcal{X}_\mathcal{D}=\{\mathcal{X}_{SL},\mathcal{X}_{TU}\}$. We then compute logits $\mathbf{y}_{SL}^{\prime}$, $\mathbf{y}_{SL}^{\mydprime}$ and $\mathbf{y}_{TU}$ as follows:
\begin{equation}
\begin{aligned}
\label{DA_Logits}
\{\mathbf{y}_{SL}^\prime,\mathbf{y}_{TU}\}&=g_{\text{MLC}}(g_{\text{SH}}(W_{\text{SH}},\mathcal{X}_\mathcal{D})),\\
\mathbf{y}_{SL}^{\mydprime}&=g_{\text{MLC}}(g_{\text{SH}}(W_{\text{SH}},\mathcal{X}_{SL})),\\
\end{aligned}
\end{equation}
notably, we utilize batch normalization as the adapter to the target distribution. Thus, only batch normalization is updated at the second forward propagation. We randomly interpolating the logits $\mathbf{y}_{SL}^\prime$ and $\mathbf{y}_{SL}^{\mydprime}$ to obtain source-domain logits $\mathbf{y}_{SL}$ as follow, 
\begin{equation}
\begin{aligned}
\label{Randomly_Interpolating}
\mathbf{y}_{SL}&=\lambda_{\text{RI}}\cdot\mathbf{y}_{SL}^\prime+(1-\lambda_{\text{RI}})\cdot\mathbf{y}_{SL}^{\mydprime},\\
\lambda_{\text{RI}}&=\text{Concat}\{\lambda_{\text{RI}_{\text{ENV}}},\lambda_{\text{RI}_{\text{DP}}}\}\in\mathbb{R}^{n_{SL}\times k},\\
\end{aligned}
\end{equation}
where $\lambda_{\text{RI}_{\text{ENV}}}\sim\mathcal{U}^{n_{SL}\cdot2}(0,1)$ and $\lambda_{\text{RI}_{\text{DP}}}\sim\mathcal{U}^{n_{SL}\cdot5}(0,1)$ for channel environment and CDL delay profiles. This formulation implicitly does $\underset{W_\text{MLC}}\min\left|\mathbf{y}_{SL}^\prime-\mathbf{y}_{SL}^{\mydprime}\right|$. We approximate the unknown $\mathbb{P}_{TU}$ using the closest available distribution $\mathbb{P}_{SL}$ that aligns the statistical property of $\mathbf{y}_{TU}=\{\mathbf{y}_{TU_{\text{DP}}},\mathbf{y}_{TU_{\text{ENV}}}\}$ with $\mathbf{y}_{SL}=\{\mathbf{y}_{SL_{\text{DP}}},\mathbf{y}_{SL_{\text{ENV}}}\}$ that helps significantly with the discriminative effectiveness \cite{AdaMatch}.

Then, we compute pseudo labels with softmax activation $\widehat{\mathbf{y}}_{SL_{\text{DP}}}$, $\widehat{\mathbf{y}}_{SL_{\text{ENV}}}$ and $\widehat{\mathbf{y}}_{TU_{\text{DP}}}$, $\widehat{\mathbf{y}}_{TU_{\text{ENV}}}$ separately. The distribution alignment rectifies the target unlabeled pseudo-labels by multiplying with the ratio of the expected value to obtain the final pseudo-labels ${\widetilde{\mathbf{y}}}_{TU}\in\mathbb{R}^{n_{TU}\times 7}$,
\begin{equation}
\begin{aligned}
\label{DistributuinA}
{\widetilde{\mathbf{y}}_{TU_{\text{DP}}}}=\text{normalize}({\widehat{\mathbf{y}}_{TU_{\text{DP}}}}\cdot\frac{\mathbb{E}\left[{\widehat{\mathbf{y}}}_{SL_{\text{DP}}}\right]}{\mathbb{E}\left[{\widehat{\mathbf{y}}}_{TU_{\text{DP}}}\right]}),\\
{\widetilde{\mathbf{y}}_{TU_{\text{ENV}}}}=\text{normalize}({\widehat{\mathbf{y}}_{TU_{\text{ENV}}}}\cdot\frac{\mathbb{E}\left[{\widehat{\mathbf{y}}}_{SL_{\text{ENV}}}\right]}{\mathbb{E}\left[{\widehat{\mathbf{y}}}_{TU_{\text{ENV}}}\right]}),
\end{aligned}
\end{equation}
the $\text{normalize}(\cdot)$ ensures that the logits sums to 1. As could be seen along the optimization, $\mathbb{E}\left[{\widetilde{\mathbf{y}}}_{TU}\right]=\mathbb{E}\left[{\widehat{\mathbf{y}}}_{SL}\right]$ confirms that distribution alignment makes the target pseudo-labels follow the source label distribution.

To selectively utilize the aligned target pseudo logits, a multi-label confidence threshold on user-provided value $\tau_\text{CT}$ is proposed. The relative confidence threshold $c_{\tau_{\text{CT}}}$ and binary mask $\mathbf{v}\in\left\{0,1\right\}^{n_{TU}}$ defined as follows,
\begin{equation}
\begin{aligned}
\label{Cthreshold}
c_{\tau_{\text{CT}}}=\frac{\tau}{n_{SL}}\sum_{i=1}^{n_{SL}}\text{Concat}\{\underset {j \in\left[1,...,5\right]} {\text{max}}{({\widehat{\mathbf{y}}}_{SL_{DP}}^{\left(i,j\right)})},\underset{j\in\left[1,2\right]} {\text{max}}({{\widehat{\mathbf{y}}}_{SL_{ENV}}^{\left(i,j\right)})}\},\\
\mathbf{v}^{(i)}=\text{Concat}\{\underset{j\in\left[1,...,5\right]} {\text{max}}{({\widetilde{\mathbf{y}}}_{TU_{DP}}^{\left(i,j\right)})},\underset{j\in\left[1,2\right]} {\text{max}}({{\widetilde{\mathbf{y}}}_{TU_{ENV}}^{\left(i,j\right)})}\}\geq c_{\tau_{\text{CT}}}.
\end{aligned}
\end{equation}

The loss function $\mathcal{L}_{\text{DA}}$ sums $\mathcal{L}_{\text{DA}_{SL}}$ for the source and $\mathcal{L}_{\text{DA}_{TU}}$ for the target.
\begin{equation}
\begin{aligned}
\label{DA_Formula}
&\mathcal{L}_{\text{DA}_{SL}}=\frac{1}{n_{SL}}\sum_{i=1}^{n_{SL}}\ell_c(\widehat{\mathbf{y}}_{SL}^{\left(i\right)},\mathbf{y}_{SL}^{\left(i\right)}),\\
&\mathcal{L}_{\text{DA}_{TU}}=\frac{1}{n_{TU}}\sum_{i=1}^{n_{TU}}\ell_c(\text{Eval}(\widetilde{\mathbf{y}}_{TU}^{\left(i\right)}),\mathbf{y}_{TU}^{\left(i\right)}))\cdot\mathbf{v}^{(i)},\\
&\mathcal{L}_{\text{DA}}(W_{\text{C}})=\mathcal{L}_{\text{DA}_{SL}}+\mu(t)\mathcal{L}_{\text{DA}_{TU}},
\end{aligned}
\end{equation}
where $\text{Eval}(\cdot)$ is the evaluation mode of model. The $\mu(t)$ is a warmup function that controls $\mathcal{L}_{\text{DA}_{TU}}$ at every step of the training, which smoothly raises from zero to one for the first half of the training and remains at one for the second half \cite{AdaMatch}.

\subsection{CSI Feature-space Adversarial Generative Stimulation}
GAN-based image compression system synthesizes details in the limit-bandwidth scenario, obtaining visually pleasing results and showing strong artefacts \cite{CGAN} \cite{Feature_DA}. The HybridCVLNet adopts a lightweight discriminator and introduces the adversarial term to the DA stage to stimulate the adaptation to the target domain in the CSI feature space.

Given a dataset $\mathcal{D}_{\mathbf{x}}$, original GANs learn to approximate distribution $\mathbb{P}_\mathbf{x}$ through a generator $G\left(\mathbf{z}\right)$ that maps samples $\mathbf{z}$ from a prior distribution $\mathbb{P}_z$ to the data distribution $\mathbb{P}_\mathbf{x}$. The generator $G$ is trained in parallel with a discriminator $Dis$ with the objective as follows,
\begin{equation}
\label{GAN_System}
\mathcal{L}_{\text{GAN}}\coloneqq\underset{Dis}{\text{max}}{\mathbb{E}\left[f(Dis(\mathbf{x}))\right]+\mathbb{E}\left[g(Dis(G(\mathbf{z})))\right]},
\end{equation}
where $f$ and $g$ are scalar functions to solve $\underset{G}{\text{min}}{\mathcal{L}_{\text{GAN}}}$ that allows to minimize general $f$-divergences between the distribution of $G(\mathbf{z})$ and $\mathbb{P}_{\mathbf{x}}$. The compression GANs (CG) for CSI feedback \cite{GAN_CSI} can be viewed as a combination of GANs and deep compression. Given latent $\mathbf{s}=E\left(\mathbf{H}\right)$, where encoder $E=\{g_{\text{SH}}, g_{\text{SH}_{\text{D}}}\circ g_{\text{UH}}, g_{\text{UH}_{\text{D}}}\}$, the saddle-point objective for GAN-based CSI feedback is formulated as,
\begin{multline}
\label{Compress_GAN}
\mathcal{L}_{\text{CG}}\coloneqq\underset{E,G}{\text{min}}\underset{Dis}{\text{max}}{{\lambda_{\text{adv}}\mathbb{E}[f(Dis(\widehat{\mathbf{H}}))]+\mathbb{E}[g(Dis(G(\mathbf{s})))]}}+\\\mathbb{E}[d(\mathbf{H},G(\mathbf{s}))],
\end{multline}
where $G$ is the Generator, $G =\{g_{\text{GT}}, g_{\text{GT}_{\text{D}}}\circ g_{\text{RT}}, g_{\text{RT}_{\text{D}}}\circ g_{\text{RB}}\}$, $\lambda_{\text{adv}}$ is the hyperparameter determines the balance between the generation and reconstruction and $d$ measures reconstruct similarity (i.e., $\ell_r$).

In domain adaptation, given source image and label pairs $(\mathcal{X}_{SL},\mathcal{Y}_{SL})\sim\mathbb{P}_{SL}$ and target images $\mathcal{X}_{TU}\sim\mathbb{P}_{TU}$. In a typical unsupervised domain adaptation (UDA) scheme that learns the source and target mapping $f_{SL}$ and $f_{TU}$ to minimize the difference of feature $f_{SL}(\mathcal{X}_{SL})$ and $f_{TU}(\mathcal{X}_{TU})$ in processing by the discriminate of the domain discriminator $Dis$. Substantially different, the primary goal of the CSI feedback is to achieve CSI regression ($f_r$). However, existing UDA schemes focus on intermediate features $\mathbf{M}_{SL}^I$, $\mathbf{M}_{TU}^I$ may not contribute. Thus a CSI feature-space stimulation DA is proposed to adapt with the $\mathbb{P}_{TU}$ as,
\begin{multline}
\label{ADV_Formula}
\mathcal{L}_{\text{Fadv}}\coloneqq\underset{E,G}{\text{min}}\underset{Dis}{\text{max}}{{\lambda_{\text{adv}}\mathbb{E}[f_{\text{LSGAN}}(Dis(\widehat{\mathbf{H}}_{TU}))]}}+\\\mathbb{E}[g_{\text{LSGAN}}(Dis(G(\mathbf{s}_{TU})))]+\mathbb{E}[d(\mathbf{H}_{TU},G(\mathbf{s}_{TU}))],
\end{multline}
where $f_{\text{LSGAN}}(\mathbf{y})={(\mathbf{y}-1)}^2$ and $g_{\text{LSGAN}}\left(\mathbf{y}\right)=\mathbf{y}^2$ (which corresponds to the Pearson $\mathcal{X}^2$ divergence) in \cite{LSGAN} to overcome the vanishing gradients problem of regular GANs, $\mathbf{s}_{TU}$ is the codeword of target domain sample. A verified CDB with anti-alias downsampling (max) and an anti-alias downsampling (average) branch \cite{anti-aliasing} is adopted as the $Dis$.

\begin{figure}[!h]
    \centering
    \includegraphics[width=2.35 in]{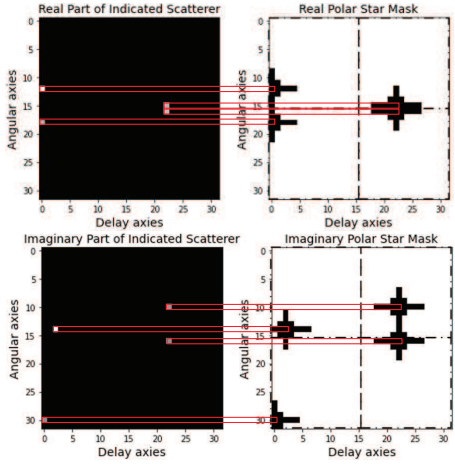}
     \caption{Illustration of a reported scatterer indication and its correspondence Polar Star Binary Mask, The red box indicates the $\text{argmax}(i_{\text{max},\bm{q}},j_{\text{max},\bm{q}})$ and the corresponding central elements of $\mathbf{M}_{\text{polar}}$.}
    \label{Polar}
    \setlength{\belowcaptionskip}{-5pt}
\end{figure}
Furthermore, inspired by the SOTA CG with a semantic mask that could fully synthesize unimportant regions \cite{DeepCompress2}. innovatively, a polar-star-like binary mask $\mathbf{M}_{\text{polar}}$ that indicates the emphasis region to synthesize is proposed. It firstly masks the $\mathbf{d}_{\text{SI}}$ sequence, and considering the amplitude fluctuation around the scatterer location is critical in final regression, a square mask $\mathbf{M}_{\text{square}}\in\mathbb{R}^{3\times3}$ is assigned. Meanwhile, a delay-direction and angular-direction masks $\mathbf{M}_{\text{delay}}\in\mathbb{R}^{1\times9}$ and $\mathbf{M}_{\text{angular}}\in\mathbb{R}^{7\times1}$ are appended to the $\mathbf{M}_{\text{polar}}=\left\{\mathbf{M}_{\text{square}},\mathbf{M}_{\text{delay}},\mathbf{M}_{\text{angular}}\right\}$. The masked GT block feature $\mathbf{M}_{{\text{GT}}}=\mathbf{M}_{\text{polar}}\odot\mathbf{M}_{\text{GT}}$ of HybridCVLNet assigns the regions of zeros corresponding to the area that should be synthesized, and regions of ones should be preserved to promote the masked region adaptation of each CSI sample. We believe such a semantic mask can facilitate adaptation to the CSI amplitude fluctuation pattern of the target domain. 

\subsection{Hybrid Feature and Distribution Domain Adaptation}
\begin{figure}[!htp]
    \centering
    \subfloat[Inductive-based transfer learning.]{
    \includegraphics[width=1.25 in]{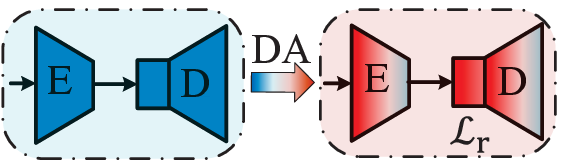}}
\quad
    \subfloat[Transductive-based domain adaptation ablation analysis.]{
    \includegraphics[width=2.55 in]{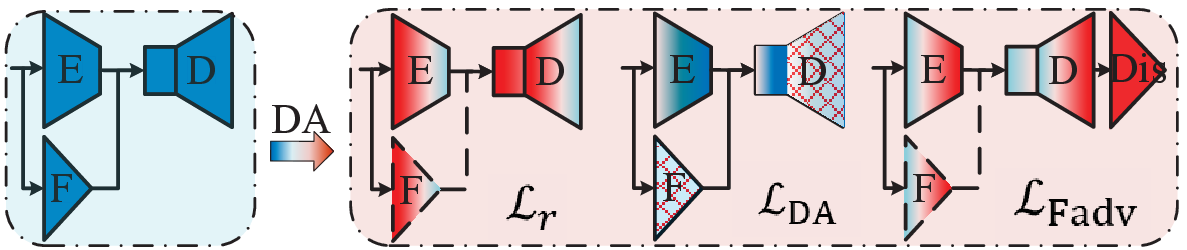}}
\quad
    \subfloat[Transductive-based hybrid domain adaptation.]{
    \includegraphics[width=1.45 in]{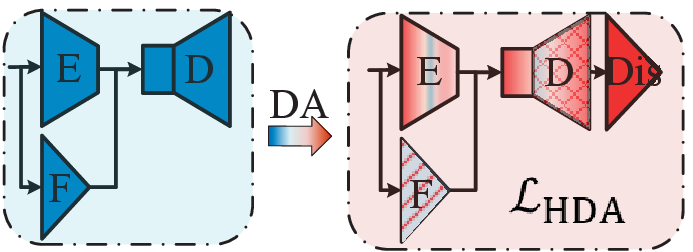}}
\caption{Domain adaptation ablation analysis, where the blue parts on the left side are the source domain model, and the red on the right side are the model adapted to the target domain. F with dashed contours under $\mathcal{L}_r$ and $\mathcal{L}_{\text{Fadv}}$ is used to denote the invalided regularization function.}
\label{DA_Objective}
\end{figure}
\vspace{-0.08cm}

The objective of HybridCLVNet HDA is a three parts combination, namely the primary target regression objective $\mathcal{L}_{\text{r}}$ from (\ref{HybridCVLNet_Objective}), the distribution-alignment objective $\mathcal{L}_{\text{DA}}$ from ML-AdaMatch loss formula (\ref{DA_Formula}), and the adversarial generative objective $\mathcal{L}_{\text{adv}}$ in (\ref{ADV_Formula}). We define three hyperparameters $\lambda_\text{Content}$, $\lambda_{\text{Rug}_{\text{HDA}}}$, and $\lambda_{\text{adv}}$ to weight each term respectively and the total learning objective is defined as follows,
\begin{multline}
\label{HDA_Loss}
\mathcal{L}_{\text{HDA}}=\lambda_\text{Content}\cdot[\mathcal{L}_{\text{r}}+\mathcal{L}_{\text{Aug}}\cdot\delta(t-\bm{Epoch}_{\text{Aug}})]+\\\lambda_{\text{Rug}_{\text{HDA}}}\cdot\mathcal{L}_{\text{DA}}+\lambda_{\text{adv}}\cdot\mathcal{L}_{\text{Fadv}}.
\end{multline}

An inductive-based adaptation to the target domain of a Vanilla CSI feedback system is shown in Fig. \ref{DA_Objective}(a), which gradually adapted with $\mathcal{L}_{\text{r}}$, yet the source domain inference performance may have an unavoidable degradation, namely the "catastrophic forgetting". Similarly, the HybridCVLNet adapted to the target domain with only $\mathcal{L}_{\text{r}}$ would forfeit prior regularized distribution knowledge, as shown in Fig. \ref{DA_Objective}(b) that ruins the category-related functions ($f_c$ and $f_s$) of HybridCVLNet. To preserve the prior regularization, we introduce the $\mathcal{L}_{\text{DA}}$ of (\ref{DA_Formula}), as shown in the central of Fig. \ref{DA_Objective}(b), and from the view of feature, the adversarial generative loss stimulation $\mathcal{L}_{\text{Fadv}}$ in (\ref{ADV_Formula}) is leveraged as shown in the right of Fig. \ref{DA_Objective}(b) to mimic the patterns of the target domain in artefact space.

The optimization strategy of HybridCVLNet HDA is to promote primary target regression optimization and preserve the prior classification latent simultaneously, as shown in Fig. \ref{DA_Objective}(c). To simplify the hyperparameters, we set the $\lambda_\text{Content}=1$, and adjust the balance between the adversarial generative $\mathcal{L}_{\text{Fadv}}$ and distribution alignment terms $\mathcal{L}_{\text{DA}}$ with two ratio factors $\lambda_{\text{Rug}_{\text{HDA}}}$ and $\lambda_{\text{adv}}$. Theoretically, the  $\mathcal{L}_{\text{DA}}$ is more effective when target and source category distribution are similar. Increasing $\lambda_{\text{Rug}_{\text{HDA}}}$ maintains the category-related performance. In comparison, increasing $\lambda_{\text{adv}}$ motivates the feature-space regression to the target domain.
\vspace{-0.15cm}
\begin{table*}[h!]
\caption{Total parameters, NMSE in $\mathrm{dB}$ and Classification precision of Intra-domain heterogeneous experiment.}
\label{Exp1}
\centering
\resizebox{0.60\linewidth}{!}{
\begin{tabular}{c|cccc|cccc} 
\hline
\multicolumn{1}{c}{CR($\gamma$)}                       & \multicolumn{4}{c}{1/32}                                                                   & \multicolumn{4}{c}{1/16}                                                                                                           \\ 
\hline\hline
\multicolumn{1}{c}{\multirow{2}{*}{Methods}} & Parameters       & NMSE~           & \multicolumn{2}{c}{Accuracy}                          & Parameters       & NMSE            & \multicolumn{2}{c}{Accuracy}                  \\
\multicolumn{1}{c}{}                         & (Total)          & (in dB)         & OF1                       & \multicolumn{1}{c}{CF1}   & (Total)          & (in dB)         & OF1                       & \multicolumn{1}{c}{CF1}    \\ 
\hline\hline
CsiNet                                       & 267,614             & -9.323          & \textbackslash{}          & \textbackslash{}          & 529,822             & -12.86          & \textbackslash{}          & \textbackslash{}         \\
CRNet                                        & 267,382          & -9.308          & \textbackslash{}          & \textbackslash{}          & 529,590          & -13.06          & \textbackslash{}          & \textbackslash{}           \\
CLNet                                        & \textbf{\textit{266,502}} & -8.066          & \textbackslash{}          & \textbackslash{}          & 528,710          & -13.35          & \textbackslash{}          & \textbackslash{}           \\
CVLNet                                       & 591,579          & \textbf{\textit{-11.83}}          & \textbackslash{}          & \textbackslash{}          & 853,787          & -13.96          & \textbackslash{}          & \textbackslash{}           \\
HybridCVLNet                                 & 311,964          & -11.59 & \textbf{\textit{98.96\%}} & \textbf{\textit{98.96\%}} & \textbf{\textit{361,308}} & \textbf{\textit{-14.24}} & \textbf{\textit{98.95}}\% & \textbf{\textit{98.96}}\%  \\
\hline
\multicolumn{1}{c}{CR($\gamma$)}                       & \multicolumn{4}{c}{1/8}                                                                   & \multicolumn{4}{c}{1/4}                                                                                                           \\ 
\hline\hline
\multicolumn{1}{c}{\multirow{2}{*}{Methods}} & Parameters       & NMSE~           & \multicolumn{2}{c}{Accuracy}                          & Parameters       & NMSE            & \multicolumn{2}{c}{Accuracy}                  \\
\multicolumn{1}{c}{}                         & (Total)          & (in dB)         & OF1                       & \multicolumn{1}{c}{CF1}   & (Total)          & (in dB)         & OF1                       & \multicolumn{1}{c}{CF1}    \\ 
\hline\hline
CsiNet                                      & 1,054,238            & -15.19          & \textbackslash{}          & \textbackslash{}          & 2,103,070            & -18.10          & \textbackslash{}          & \textbackslash{}          \\
CRNet                           & 1,054,006        & -15.50          & \textbackslash{}          & \textbackslash{}          & 2,102,838        & -15.96          & \textbackslash{}          & \textbackslash{}           \\
CLNet                      & 1,053,126        & -15.95          & \textbackslash{}          & \textbackslash{}          & 2,101,958        & -18.44          & \textbackslash{}          & \textbackslash{}           \\
CVLNet            & 1,378,203        & -15.31          & \textbackslash{}          & \textbackslash{}          & 2,427,035        & -18.32          & \textbackslash{}          & \textbackslash{}          \\
HybridCVLNet           & \textbf{\textit{459,996}} & \textbf{\textit{-16.65}} & \textbf{\textit{98.95}}\% & \textbf{\textit{98.96}}\% & \textbf{\textit{657,372}} & \textbf{\textit{-19.25}} & \textbf{\textit{98.95}}\%  & \textbf{\textit{98.96}}\%   \\
\hline
\end{tabular}
}
\end{table*}

\begin{table}[h]
\captionsetup{size=footnotesize}
\caption{nrCDLChannel Model Parameters.} \label{CDL_param}
\setlength\tabcolsep{0pt} 
\footnotesize\centering

\smallskip 
\resizebox{0.55\linewidth}{!}{
\begin{tabular*}{0.65\columnwidth}{@{\extracolsep{\fill}}cccccccc}
\toprule
\multicolumn{4}{c}{Parameters}    &\multicolumn{4}{c}{Values} \\
\midrule
  \multicolumn{4}{c}{NR Resource Block (RB)} & \multicolumn{4}{c}{51} \\
  \multicolumn{4}{c}{Subcarriers} & \multicolumn{4}{c}{612 (12 per RB)} \\
  \multicolumn{4}{c}{Subcarrier Spacing} & \multicolumn{4}{c}{30 kHz} \\
  \multicolumn{4}{c}{Carrier Frequency} & \multicolumn{4}{c}{3.5 GHz} \\
  \multicolumn{4}{c}{BS Antennas} & \multicolumn{4}{c}{32 Monopolar ULA} \\
  \multicolumn{4}{c}{UE Antenna} & \multicolumn{4}{c}{single-antenna} \\
  \multicolumn{4}{c}{UE Speed} & \multicolumn{4}{c}{30 km/h} \\
\bottomrule
\end{tabular*}
}
\end{table}  

\section{Simulation Results And Analysis}
In this section, we conduct two sets of experiments to validate and analyze the performance and generalizability of the proposed HybridCVLNet in an intra-domain heterogeneous dataset and the implementation and robustness of the transductive-based hybrid domain adaptation frameworks under two inter-domain data drift settings, respectively.
\vspace{-0.35cm}

\subsection{Intra-domain Heterogeneous Experiment}
In this experiment, we evaluate the performance of the HybridCVLNet with its learning objective on the heterogeneous regularization dataset. As the theoretical design, the proposed HybridCVLNet can effectively preserve the category-related features and fuse with category-invariant features for refinement. The challenge is to obtain guaranteed regression manifestation (achieves the $f_r$) and high classification accuracy (achieves the $f_c$) simultaneously.

The dataset $\mathcal{D}_{SL}$ is generated on the CDL channel model described in Section IV with parameter settings shown in Table \ref{CDL_param} in a perfect downlink channel estimation mode to obtain $\mathbf{H}\in\mathbb{C}^{2\times32\times612}$. Latterly was processed with domain transformation to ADCRM, where the angular resolution sampling point is chosen as the number of antennas, and the delay sample point is 400 to obtain $\mathbf{H}_{\text{AD}}\in\mathbb{C}^{2\times32\times400}$. Considering the channel sparsity in the large delay range, the $\mathbf{H}_{\text{AD}}$ is truncated to 32 along the delay axis that $\mathbf{H}_{\text{AD}}^\prime\in\mathbb{C}^{2\times32\times32}$, and is normalized with the mean value of $0.5$ by min-max normalization. The $\mathcal{D}_{SL}$ randomly shuffled five types of CDL delay profiles from CDL-A to E, and $30\%$ of the category-balanced dataset is chosen for testing ($N_{\text{train}}=112,000$ and $N_{\text{test}}=48,000$).  

\begin{table*}[h!]
\caption{NMSE, GGAP in $\mathrm{dB}$ and Classification precision of HybridCVLNet Ablation in Intra-domain heterogeneous Dataset.}
\label{Exp1_AB}
\centering
\resizebox{0.65\linewidth}{!}{
\begin{tabular}{c|cccc|cccc} 
\hline
\multicolumn{1}{c}{CR($\gamma$)}                       & \multicolumn{4}{c}{1/32}                                                                   & \multicolumn{4}{c}{1/16}                                                                                                           \\ 
\hline\hline
\multicolumn{1}{c}{\multirow{2}{*}{Ablation Settings}} &\multicolumn{3}{c}{NMSE (in dB)}          & \multicolumn{1}{c}{Accuracy}          &\multicolumn{3}{c}{NMSE (in dB)}            & Accuracy                  \\
\multicolumn{1}{c}{}                         & Train          & Test         & $\text{GGAP}_{\mathcal{D}_{SL}}$                       & \multicolumn{1}{c}{OF1}   & Train          & Test        & $\text{GGAP}_{\mathcal{D}_{SL}}$                       & \multicolumn{1}{c}{OF1}    \\ 
\hline\hline
HybridCVLNet(Plain)                                       & -11.01             & -7.768          & 3.24          & \textbackslash{}            & -13.79             & -8.897         & 4.89          & \textbackslash{}           \\
HybridCVLNet(Aug)                                       & -10.81          & -10.10          & 0.71          & \textbackslash{}           & -13.30         & -11.96          & 1.34          & \textbackslash{}           \\
HybridCVLNet(Reg)                                        & -11.40          & -10.72          & 0.68          & 98.94\%          & -12.88          & -12.03          & 0.85         & 98.44\%           \\
\hline
\multicolumn{1}{c}{CR($\gamma$)}                       & \multicolumn{4}{c}{1/8}                                                                   & \multicolumn{4}{c}{1/4}                                                                                                           \\ 
\hline\hline
\multicolumn{1}{c}{\multirow{2}{*}{Ablation Settings}}  &\multicolumn{3}{c}{NMSE (in dB)}          & \multicolumn{1}{c}{Accuracy}          &\multicolumn{3}{c}{NMSE (in dB)}            & Accuracy                  \\
\multicolumn{1}{c}{}                         & Train          & Test         & $\text{GGAP}_{\mathcal{D}_{SL}}$                       &\multicolumn{1}{c}{OF1}   & Train          & Test        &$\text{GGAP}_{\mathcal{D}_{SL}}$                       & \multicolumn{1}{c}{OF1}    \\ 
\hline\hline
HybridCVLNet(Plain)                                      & -16.13            & -11.71         & 4.42         &\textbackslash{}           & -18.50            & -13.87          & 4.63         &\textbackslash{}         \\
HybridCVLNet(Aug)                       & -15.48        & -14.47          & 1.01         &\textbackslash{}          & -17.92        & -16.86          & 0.96         & \textbackslash{}  \\
HybridCVLNet(Reg)                            & -14.89        & -14.52          & 0.37          & 98.96\%           & -17.82        & -17.16          &0.66         & 98.96\%          \\
\hline
\end{tabular}
 }
\end{table*}
Several Key Performance Indicators (KPIs) are considered for the thorough assessment. The regression task $f_r$ of HybridCVLNet conducts the global normalized mean squared error (NMSE) as the intra-domain heterogeneous CSI reconstruction quality metric. The regularization task $f_c$ adopts the overall and per-category F1-measure (OF1 and CF1) to evaluate the classification accuracy. The expression of NMSE in decibels (dB) is given by:
\begin{equation}
\label{NMSE}
\text{NMSE}=10\text{lg}\mathbb{E}\{\frac{\Vert\mathbf{H}_{\text{AD}}^\prime-\widehat{\mathbf{H}}_{\text{AD}}^\prime\Vert_2^2}{\Vert\mathbf{H}_{\text{AD}}^\prime\Vert_2^2}\},
\end{equation}
alternatively, the OF1 and CF1 between the final logits $\widehat{\mathbf{y}}$ and the target label $\mathbf{y}$ defined as
\begin{equation}
\begin{aligned}
\label{Experiment1_Metrics}
\text{OF1}&=\frac{2\times\text{Precision}\times\text{Recall}}{\text{Precision}+\text{Recall}},\\
\text{CF1}&=\frac{1}{k}\sum_{i=1}^{k}\frac{2\times\text{Precision}_i\times\text{Recall}_i}{\text{Precision}_i+\text{Recall}_i}, \text{where}\\
\text{Precision}_i&=\frac{\text{TruePositive}(\widehat{\mathbf{y}}^i,\mathbf{y}^i)}{\text{FalsePositive}(\widehat{\mathbf{y}}^i,\mathbf{y}^i)+\text{TruePositive}(\widehat{\mathbf{y}}^i,\mathbf{y}^i)},\\
\text{Recall}_i&=\frac{\text{TruePositive}(\widehat{\mathbf{y}}^i,\mathbf{y}^i)}{\text{FalsePositive}(\widehat{\mathbf{y}}^i,\mathbf{y}^i)+\text{FalseNegative}(\widehat{\mathbf{y}}^i,\mathbf{y}^i)}.\\
\end{aligned}
\end{equation}

To be fair in revealing the performance of the proposed HybridCVLNet, based on the discussion of the critical factors of deep learning generalizability, we list the performance comparison among original benchmark CsiNet \cite{CsiNet} that balanced between performance and parameters, the SOTA performance-seeking CRNet \cite{CRNet}, the benchmark lightweight CLNet \cite{CLNet} with extreme low parameter and complexity, and the SOTA complex-valued CVLNet \cite{CVLNet}. In particular, to the best of our knowledge, there is no related literature on CSI feedback that utilizes discriminative tasks to regularize. The PRVNet \cite{VAE_CSI}, as a VAE CSI feedback scheme, is considered an effective partial regularity scheme that uses the KL-divergence loss term. Nevertheless, it is not convergent in the heterogeneous dataset $\mathcal{D}_{SL}$.

The HybridCVLNet follows a balanced learning strategy and sets the hyperparameters $\lambda_\text{Rug}$ of the objective function \eqref{HybridCVLNet_Objective} to $1e^{-3}$. The feature-augmentation is involved when the discriminative $\mathcal{L}_\text{c}$ and regression $\mathcal{L}_\text{r}$ terms of \eqref{HybridCVLNet_Objective} are convergence smoothly. Thus we set the $\bm{Epoch}_{\text{Ar}}\in{\{70,...,150\}}$. 

Table \ref{Exp1} summarizes the overall performance comparison. The network size is measured by total (UE and BS side) parameter numbers directly indicating the model size. The HybridCVLNet can achieve a significant parameter reduction of about $72.9\%$ when $\gamma=1/4$, $66.6\%$ when $\gamma=1/8$, $57.7\%$ and $47.2\%$ when $\gamma=1/16$ and $\gamma=1/32$ with a stable advantage on inference, respectively. Our parametric advantage comes mainly from the CDB and UDB that process features while mapping. In addition, lightweight design throughout the HybridCVLNet, such as depthwise separable convolution and layer normalization from RFA, the inverse bottleneck structure. We also calculate the network complexity with the floating-point operations per second (FLOPs) of the HybridCVLNet, which is $26.3$M when $\gamma=1/16$ that is competitive with SOTA lightweight CSI feedback system \cite{ShuffleNet} with $24.3$M of UE side. We believe that the lightweight structure of HybridCVLNet offers the possibility of improving the generalizability of the system, which is also in line with the conventional neural network generalization theory.

\begin{figure*}[ht]
    \centering
    \includegraphics[width=6.20 in]{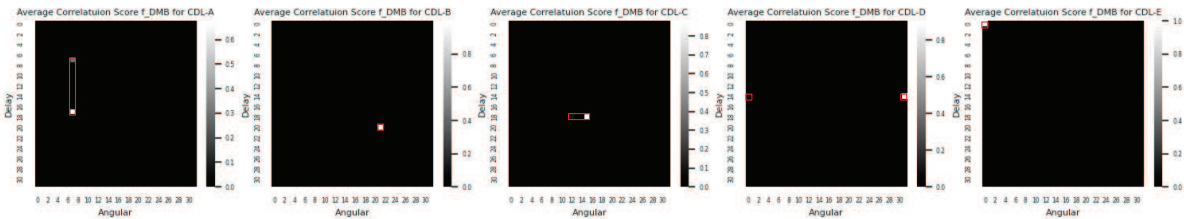}
     \caption{Enumeration of the average $\mathbf{f}_{\text{DMB}}$ in (\ref{Distribution_Mapping}) under CDL-A/B/C/D/E channel profiles. The non-zero values of $\mathbf{f}_{\text{DMB}}$ are sparse due to the use of the sharper Softmax as activation. We observe that the non-zero values of $\mathbf{f}_{\text{DMB}}$ for CDL-D and CDL-E representing LOS scenarios have lower delays compared to the $\mathbf{f}_{\text{DMB}}$ for CDL-A/B/C representing NLOS scenarios, which is in line with the perception for multipath channel propagation delays.}
    \label{fDMB}
    \setlength{\belowcaptionskip}{-5pt}
\end{figure*}

As a game between performance-seeking and generalizability, our proposed HybridCVLNet retains the high accuracy required for CSI reporting while achieving high discriminative precision. Specifically, the preliminary reconstruction inference results show that HybridCVLNet achieves a stable gain of about 1 dB compared to other comparative schemes under four different compression rates. The accuracy of OF1 and CF1 remains around $98.96\%$ independent of the compression rate. The decoder for HybridCVLNet makes full use of the stochastic codeword $\mathbf{d}_{\text{CL}}$ and considers an implicit mapping function $f_s$ to map statistical reporting to enhance the feature-level reconstruction $f_r$. To verify the validity of guidance from $f_s$, we demonstrate the average logits correlation scores $\mathbf{f}_{\text{DMB}}$ from the formula (\ref{Distribution_Mapping}) for CDL-A/B/C/D/E respectively, as shown in Fig. \ref{fDMB}.

To verify the contribution of each term of formula (\ref{Loss_One}) to the joint objective function (\ref{HybridCVLNet_Objective}), we enumerate the ablation results among the plain HybridCVLNet(Plain) that with only regression objective $\mathcal{L}_{\text{r}}$, the regularized plain HybridCVLNet(Reg) that without augmentation objective $\mathcal{L}_{\text{Aug}}$ and the augmented plain HybridCVLNet(Aug) that without regularization objective $\mathcal{L}_{\text{c}}$, are demonstrated in Table \ref{Exp1_AB}. To evaluate the regularization of $f_c$ to the $f_r$ of the HybridCVLNet that overcomes the dataset bias, the indicator generalizability gap (GGAP) is adopted. The GGAP is defined as the difference in CSI reconstruction performance between the training and inference stage. The expression of GGAP in decibels (dB) is given by:

\begin{equation}
\label{NMSE}
\text{GGAP}_{\mathcal{D}_{SL}}=\text{NMSE}_{{\mathcal{D}_{SL}}_{\text{test}}}-\text{NMSE}_{{\mathcal{D}_{SL}}_{\text{train}}}.
\end{equation}

The HybridCVLNet(Plain) with hybrid architecture is a preliminary version without the regularized and augmented constrained. Its maximum GGAP reaches a $4.89$ dB when $\gamma=1/16$ and has no accurate classification output. The HybridCVLNet(Aug) under augmented constrained $\mathcal{L}_{\text{Aug}}$ with the same set of  $\bm{Epoch}_{\text{Ar}}$ obtains up to $3.29$ dB advantage over the HybridCVLNet(Plain) when $\gamma=1/4$ and the GGAP value is significantly reduced. However, the regularized discriminative task remains invalid. The HybridCVLNet(Reg) with regularization $\mathcal{L}_{\text{c}}$ adapts to the hybrid structure of the HybridCVLNet, which achieves maximum inference NMSE performances and lowest GGAP values as a sign of overcoming the dataset bias among four $\gamma$ settings, and affects the optimization away from the tendency of overfitting. Moreover, the accurate classification result compensates for the regression degradation due to the implicit mapping function $f_s$ of the decoder of the HybridCVLNet.  
\vspace{-0.12cm}

\begin{table*}[ht!]
\caption{Total parameters, Target NMSE in $\mathrm{dB}$ and Source Classification precision of Inter-domain Domain Adaptation Experiment.}
\label{Exp2}
\centering
\resizebox{0.85\textwidth}{!}{
\begin{tabular}{cc|ccccc|ccccc}
\hline
\multicolumn{1}{c}{CR($\gamma$)}                       & \multicolumn{11}{c}{1/4}                                                                 \\
\hline\hline
\multicolumn{1}{c}{\multirow{2}{*}{Methods}} & \multicolumn{1}{c}{\multirow{2}{*}{Parameters}}  & \multicolumn{3}{c}{$\mathcal{D}_{TU1}$  $\text{NMSE}_{\text{ER}}$ (in dB)}      & Target $\mathcal{D}_{TU1}$      & \multicolumn{1}{c}{Source $\mathcal{D}_{SL}$}   & \multicolumn{3}{c}{$\mathcal{D}_{TU2}$ NMSE (in dB)}  & Target $\mathcal{D}_{TU2}$      & \multicolumn{1}{c}{Source $\mathcal{D}_{SL}$} \\
&\multicolumn{1}{c}{}     & Train      & Test     & $\text{GGAP}_{\text{ER}}$    & $\rho$   &\multicolumn{1}{c}{OF1}      & Train         & Test     & GGAP    & $\rho$  &  \multicolumn{1}{c}{OF1}\\
\hline\hline
HybridCVLNet(DT)                                      & \textbf{\textit{657,372}}      & -5.197   & -4.216  & 0.981   & 0.56          & 91.50\%    & -8.450   & -8.441  & \textbf{\textit{0.009}}   & 0.71          & 95.90\% \\
HybridCVLNet($\text{DA}_{D}$)                                      & \textbf{\textit{657,372}}            & -5.844   & -4.810  & 1.034   & 0.58          & \textbf{\textit{96.30}}\%              & -10.82   & -9.263  & 1.557   & 0.72          & 78.25\%    \\
HybridCVLNet($\text{DA}_{F}$)                                      & 674,196            & -5.516   & -4.729  & \textbf{\textit{0.787}}   & 0.57          & 76.50\%              & -11.76   & -9.607  & 2.153   & \textbf{0.73}          & 98.16\%    \\
HybridCVLNet($\text{DA}_{H}$)                                      & 674,196            & -5.821   & \textbf{\textit{-4.801}} & 1.020   &  \textbf{\textit{0.59}}          & 91.34\%              & -11.51   & \textbf{-9.628}  & 1.882   & \textbf{0.73}          & \textbf{\textit{98.75}}\%    \\
\hdashline
CVLNet(DT)                                      & 2,427,035            & -7.395   & -4.758  & 2.637   & \textbf{\textit{0.59}}          & \textbackslash{}            & -15.52   & \textit{-12.28}  & 3.240   & \textit{0.74}          & \textbackslash{}   \\
\hline
\multicolumn{1}{c}{CR($\gamma$)}                       & \multicolumn{11}{c}{1/8}                                                                 \\
\hline\hline
\multicolumn{1}{c}{\multirow{2}{*}{Methods}} & \multicolumn{1}{c}{\multirow{2}{*}{Parameters}}  & \multicolumn{3}{c}{$\mathcal{D}_{TU1}$ $\text{NMSE}_{\text{ER}}$ (in dB)}      & Target $\mathcal{D}_{TU1}$      & \multicolumn{1}{c}{Source $\mathcal{D}_{SL}$}    & \multicolumn{3}{c}{$\mathcal{D}_{TU2}$ NMSE (in dB)}  & Target $\mathcal{D}_{TU2}$      & \multicolumn{1}{c}{Source $\mathcal{D}_{SL}$}  \\
&\multicolumn{1}{c}{}     & Train      & Test     & $\text{GGAP}_{\text{ER}}$    & $\rho$   &\multicolumn{1}{c}{OF1}      & Train         & Test     & GGAP    & $\rho$  &  \multicolumn{1}{c}{OF1}\\
\hline\hline
HybridCVLNet(DT)                                      & \textbf{\textit{459,996}}            & -3.865   & -3.156  & 0.709  & 0.53          & 97.40\%              & -7.801   & -6.636  & \textbf{\textit{1.165}}   & 0.69          & 97.40\%    \\
HybridCVLNet($\text{DA}_{D}$)                                      & \textbf{\textit{459,996}}            & -3.762   & -3.462  & \textbf{\textit{0.130}}   & 0.54          & 97.20\%              & -8.032   & -6.804  & 1.228   & 0.69          & 88.82\%    \\
HybridCVLNet($\text{DA}_{F}$)                                      & 476,820            & -3.688   & -3.430  & 0.258   & 0.54          & 80.46\%              & -8.315   & -6.892  & 1.423   & 0.69          & 71.50\%    \\
HybridCVLNet($\text{DA}_{H}$)                                      & 476,820            & -4.029   & \textbf{\textit{-3.522}}  & 0.507   & \textbf{\textit{0.55}}          & \textbf{\textit{98.17}}\%              & -8.498   & \textbf{-7.207}  & 1.291   & \textbf{0.70}          & \textbf{\textit{98.72}}\%    \\
\hdashline
CVLNet(DT)                                      & 1,378,203            & -4.067   & -2.568  & 1.499   & 0.49          & \textbackslash{}            & -11.354   & \textit{-8.927}  & 2.427   & \textit{0.72}          & \textbackslash{}   \\
\hline
\multicolumn{1}{c}{CR($\gamma$)}                       & \multicolumn{11}{c}{1/16}                                                                 \\
\hline\hline
\multicolumn{1}{c}{\multirow{2}{*}{Methods}} & \multicolumn{1}{c}{\multirow{2}{*}{Parameters}}  &\multicolumn{3}{c}{$\mathcal{D}_{TU1}$ $\text{NMSE}_{\text{ER}}$ (in dB)}     & Target $\mathcal{D}_{TU1}$      & \multicolumn{1}{c}{Source $\mathcal{D}_{SL}$}   & \multicolumn{3}{c}{$\mathcal{D}_{TU2}$ NMSE (in dB)}  & Target $\mathcal{D}_{TU2}$     & \multicolumn{1}{c}{Source $\mathcal{D}_{SL}$} \\
&\multicolumn{1}{c}{}     & Train      & Test     & $\text{GGAP}_{\text{ER}}$    & $\rho$   &\multicolumn{1}{c}{OF1}      & Train         & Test     & GGAP    & $\rho$  &  \multicolumn{1}{c}{OF1}\\
\hline\hline
HybridCVLNet(DT)                                      & \textbf{\textit{361,308}}            & -3.185   & -2.180  & 1.005   & 0.51          & 89.50\%              & -6.168   & -4.818  & 1.350   & 0.65          & 89.50\%    \\
HybridCVLNet($\text{DA}_{D}$)                                      & \textbf{\textit{361,308}}            & -2.733   & -2.232 & \textbf{\textit{0.501}}   & \textbf{\textit{0.52}}          & \textbf{\textit{97.77}}\%              & -6.566   & -5.302  & \textbf{\textit{1.264}}   & 0.67          & 89.10\%    \\
HybridCVLNet($\text{DA}_{F}$)                                      & 378,132            & -2.921   & -2.222  & 0.699   & \textbf{\textit{0.52}}          & 69.20\%              & -6.939   & -5.271  & 1.668   & 0.67          & 57.06\%    \\
HybridCVLNet($\text{DA}_{H}$)                                      & 378,132            & -3.319   & \textbf{\textit{-2.550}}  & 0.769   & \textbf{\textit{0.52}}          & 94.80\%              & -6.654   & \textbf{\textit{-5.339}}  & 1.315   & \textbf{\textit{0.68}}          & \textbf{\textit{92.40}}\%    \\
\hdashline
CVLNet(DT)                                      & 853,787            & -4.040   & -2.007  & 2.033   & 0.46          & \textbackslash{}            & -8.877   & -6.349  & 2.528   & \textbf{\textit{0.68}}          & \textbackslash{}   \\
\hline
\multicolumn{1}{c}{CR($\gamma$)}                       & \multicolumn{11}{c}{1/32}                                                                 \\
\hline\hline
\multicolumn{1}{c}{\multirow{2}{*}{Methods}} & \multicolumn{1}{c}{\multirow{2}{*}{Parameters}}  & \multicolumn{3}{c}{$\mathcal{D}_{TU1}$ $\text{NMSE}_{\text{ER}}$ (in dB)}      & Target $\mathcal{D}_{TU1}$     & \multicolumn{1}{c}{Source $\mathcal{D}_{SL}$}   & \multicolumn{3}{c}{$\mathcal{D}_{TU2}$ NMSE (in dB)}  & Target $\mathcal{D}_{TU2}$      & \multicolumn{1}{c}{Source $\mathcal{D}_{SL}$} \\
&\multicolumn{1}{c}{}     & Train      & Test     & $\text{GGAP}_{\text{ER}}$    & $\rho$   &\multicolumn{1}{c}{OF1}      & Train         & Test     & GGAP    & $\rho$  &  \multicolumn{1}{c}{OF1}\\
\hline\hline
HybridCVLNet(DT)                                      & \textbf{\textit{311,964}}            & -2.124   & -1.639  & 0.485   & 0.48          & 51.70\%              & -3.854   & -3.311  & 0.543   & 0.60          & 90.50\%    \\
HybridCVLNet($\text{DA}_{D}$)                                      & \textbf{\textit{311,964}}           & -2.651   & -2.125  & 0.526   & 0.50          & 79.10\%              & -4.314   & -3.86  & \textbf{\textit{0.454}}   & 0.62          & 85.90\%    \\
HybridCVLNet($\text{DA}_{F}$)                                      & 328,788          & -2.686   & -2.096  & 0.590   & 0.50          & 93.80\%    & -5.238   & -4.022  & 1.216   & \textbf{\textit{0.63}}          & 78.74\%  \\
HybridCVLNet($\text{DA}_{H}$)                                      & 328,788            & -2.642   & \textbf{\textit{-2.381}}  & \textbf{\textit{0.261}}   & \textbf{\textit{0.52}}          & \textbf{\textit{96.44}}\%    & -4.805   & \textbf{-4.036}  & 0.769   & \textbf{\textit{0.63}}          & \textbf{\textit{91.25}}\% \\
\hdashline
CVLNet(DT)                                      &591,579       & -2.293   & -1.315  & 0.978   & 0.40          & \textbackslash{}  & -6.897   & \textit{-4.238}  & 2.659   & \textbf{\textit{0.63}}          & \textbackslash{}    \\
\hline
\end{tabular}
}
\end{table*}

\begin{table}[ht]
\captionsetup{size=footnotesize}
\caption{Model Parameters of the Target COST2100.} \label{COST_param}
\setlength\tabcolsep{0pt} 
\footnotesize\centering

\smallskip 
\resizebox{0.90\linewidth}{!}{
\begin{tabular*}{\columnwidth}{@{\extracolsep{\fill}}cccccccccccc}
\toprule
\multicolumn{4}{c}{Parameters}    &\multicolumn{4}{c}{$\mathbb{D}^1$ Values}    &\multicolumn{4}{c}{$\mathbb{D}^2$ Values} \\
\midrule
  \multicolumn{4}{c}{Channel Environment} & \multicolumn{4}{c}{IndoorHall-5GHz} & \multicolumn{4}{c}{SemiUrban-300MHz} \\
  \multicolumn{4}{c}{Scenario} & \multicolumn{4}{c}{LOS} & \multicolumn{4}{c}{NLOS}\\
  \multicolumn{4}{c}{Frequency Band} & \multicolumn{4}{c}{5.1GHz-5.3GHz} & \multicolumn{4}{c}{275MHz-295MHz}\\
  \multicolumn{4}{c}{BS Antennas} & \multicolumn{4}{c}{32 Omni-VLA} & \multicolumn{4}{c}{32 Omni-VLA} \\
  \multicolumn{4}{c}{UE Antenna} & \multicolumn{4}{c}{single-antenna} & \multicolumn{4}{c}{single-antenna}\\
  \multicolumn{4}{c}{UE Speed} & \multicolumn{4}{c}{0.1m/s-0 m/s} & \multicolumn{4}{c}{0.2m/s-0.9 m/s}\\
\bottomrule
\end{tabular*}
}
\end{table}

\subsection{Inter-domain Domain Adaptation Experiment}
In this experiment, we will simulate the adaptation process of an online deployed CSI feedback model. In the scenario of domain adaptation applications, the size of the target data collected online is limited, and the training epochs are nonredundant due to the restricted UE-side computing overhead. We want to use this experiment to present the performance of the HybridCVLNet inductive-based fine-tuning, namely the direct transform (DT) on the target domain (i.e., whether a priori category-related knowledge can promote regression to overcome the data drift), and demonstrate the robustness and performance of the proposed transductive-based hybrid domain adaptation (DA) scheme in homogeneous and heterogeneous target channel environment datasets. Further, we will briefly explore the difference between networks pursuing generalizability and SOTA seeking dedicated and localized performance.  

The target unlabeled domain dataset $\mathcal{D}_{TU}$ is generated from the COST2100 \cite{COST2100} model, regarded as a benchmark dataset in the CSI feedback. The parameter setting of target domain one $\mathbb{D}^1$ and domain two $\mathbb{D}^2$  is listed in Table \ref{COST_param}, and process to $\mathbf{H}^\prime_{\text{AD}_{TU}}\in\mathbb{C}^{2\times32\times32}$, $\mathbf{H}^\prime_{\text{AD}_{TU}}\subset\mathcal{D}_{TU}\sim\mathbb{D}^1, \mathbb{D}^2$. We placed the $\mathcal{D}_{TU}$ into two groups. In Group one, to evaluate the adaptation performance towards the mixed-distribution target dataset, we set $\mathcal{D}_{TU1}\sim\mathbb{D}^1, \mathbb{D}^2$. The second group is to validate the performance under the homogenous dataset, where $\mathcal{D}_{TU2}\sim\mathbb{D}^1$ only. We set the accessible target data sample to 2,000 ($M_{TU1_{\text{train}}}=M_{TU2_{\text{train}}}=2,000$), which are randomly picked from the $\mathcal{D}_{TU1}$ and $\mathcal{D}_{TU2}$ each consisting of 150,000 ($M_{TU1}=M_{TU2}=150,000$) samples respectively. The training epoch of fine-tuning and domain adaptation is set to 40 rounds, $\bm{Epoch}_{\text{DT}} = \bm{Epoch}_{\text{DA}}\in{\{1,...,40\}}$.

Several KPIs of robustness are involved in validating the proposed domain adaptation framework for the HybridCVLNet. 
\begin{equation}
\begin{aligned}
\label{Experiment_Metric2}
\text{NMSE}_{\text{ER}}&=\frac{1}{N}\sum_{j=1}^{N} \text{NMSE}_{\mathbb{D}_{\text{test}}^j}(\mathbf{H}_{\text{AD}_{\text{test}}}^\prime,\widehat{\mathbf{H}}_{\text{AD}_{\text{test}}}^\prime),\\
\text{GGAP}_{\text{ER}}&=\text{NMSE}_{\text{ER}_{\text{test}}}-\text{NMSE}_{\text{ER}_{\text{train}}},\\
\rho&=\mathbb{E}\left\{\frac{1}{N_c}\sum_{\mathrm{n}=1}^{N_{c}}\frac{\left|{\hat{\mathbf{h}}^\mathrm{H}}_\mathrm{n}\mathbf{h}_\mathrm{n}\right|}{{\Vert\hat{\mathbf{h}}_\mathrm{n}\Vert_2}{\Vert \mathbf{h}_\mathrm{n}\Vert_2}}\right\},\\
\end{aligned}
\end{equation}
in which an Empirical Risk NMSE ($\text{NMSE}_{\text{ER}}$) across $N$ target domain in dB can perform as the metrics for measuring the generalizability and stability to the dataset bias of statistical learning or neural network. Alternatively, the Empirical Risk generalizability gap ($\text{GGAP}_{\text{ER}}$) is adopted. Meanwhile, for target dataset $\mathcal{D}_{TU}$ maintains the frequency channel response matrix ${\mathbf{H}}_{\text{multi}}$, the cosine similarity $\rho$ which directly compares the similarity of the predicted matrices $\widehat{\mathbf{H}}_{\mathrm{multi}}$ and $\mathbf{H}_{\mathrm{multi}}$, where ${\widehat{\mathbf{h}}}_\mathrm{n}$ and $\mathbf{h}_\mathrm{n}$ denote the channel responses of the $n$-th subcarrier.

Table \ref{Exp2} summarizes the ablation experiments performance comparison of HybridCVLNet DT mode and proposed DA scheme from distribution ($\text{DA}_D$), features ($\text{DA}_F$) and hybrid scheme ($\text{DA}_H$) with the target domain dataset $\mathcal{D}_{TU1}$ and $\mathcal{D}_{TU2}$. The performance of the SOTA network seeking dedicated performance (i.e. CVLNet) is distinguished by a dashed line. We enumerate the training parameters, where HybridCVLNet $\text{DA}_F$ and $\text{DA}_H$ must include the lightweight CDB discriminator, with a slightly parametric increment of 16,824 for all four $\gamma$ compared to DT and $\text{DA}_D$ modes. Nevertheless, it remains a significant parametric superior to the SOTA CVLNet.

Inter-domain experiment group one reveals that our proposed HybridCVLNet in both the DT and all DA modes achieves performance beyond the $\text{NMSE}_{\text{ER}}$ and similarity $\rho$ of the CVLNet DT mode under all the compression rates from $\gamma=1/32$ to $\gamma=1/4$ in more heterogeneous target dataset $\mathcal{D}_{TU1}$ randomly sampled from domain $\mathbb{D}^1$ and $\mathbb{D}^2$. Specifically, the transductive-based $\text{DA}_H$ mode of HybridCVLNet achieves a performance boost of up to $50\%$ on $\text{NMSE}_{\text{ER}}$ when $\gamma=1/32$, and achieves an up to nearly $13\%$ improvement on $\rho$ when $\gamma=1/8$ and $\gamma=1/16$. Meanwhile, it preserves the critical regularization by maintaining the overall F1-measure (OF1) of at least $91.34\%$ of classification accuracy in source domain dataset $\mathcal{D}_{SL}$. We also list the $\text{GGAP}_{\text{ER}}$ under the heterogeneous target dataset to demonstrate that HybridCVLNet and the proposed DA modes are less prone to overfitting and stabilized than the SOTA CVLNet.  

Inter-domain experiment group two presents the adaptation from the source domain to a more homogeneous target dataset $\mathcal{D}_{TU2}$ sampled from domain $\mathbb{D}^2$. Compared to the performance-seeking SOTA CVLNet, the proposed HybridCVLNet with generalizability design and miniature parameter space embodied is a recession at higher compression rates of $\gamma=1/4, 1/8, 1/16$ on NMSE but is significantly superior in cosine similarity performance $\rho$, with a slight performance gap occurring at a gamma of 1/8, reaching most 0.02. We speculate that such a similarity gain comes mainly from our hybrid codewords, especially the sample-level channel scatterer indication $\mathbf{d}_\text{SI}$. Moreover, at all compression rates, the HybridCVLNet outperforms the CVLNet and achieves an impressive 3.231 dB advantage when $\gamma=1/4$ on the generalizability measurement GGAP.

The results of the ablation experiment in $\mathcal{D}_{TU1}$ and $\mathcal{D}_{TU2}$ reveal the superior effectiveness of transductive-based $\text{DA}_F$ and $\text{DA}_D$ modes to the inductive-based DT with stable performance gains on $\text{NMSE}_{\text{ER}}$ and $\rho$. Theoretically, when $\mathcal{D}_{TU}$ is similar to the $\mathcal{D}_{SL}$ ($\mathcal{D}_{TU1}$, for instance), the performance gain of the $\text{DA}_D$ scheme will be larger than $\text{DA}_F$, as verified at different compression rates. By contrast, because of the difference in categorical distribution between the $\mathcal{D}_{TU2}$ and source dataset $\mathcal{D}_{SL}$, an arbitrary distribution alignment would invalidate the classification mapping $f_{c}$ and implicit mapping $f_{s}$, as evidenced by the results of ablation experiments, where the $\text{NMSE}$ and similarity $\rho$ of the feature adaptation scheme $\text{DA}_F$ outperformed the distribution alignment scheme $\text{DA}_D$ at multiple compression rates $\gamma$. Ultimately, through the balancing of the $\mathcal{L}_{\text{Fadv}}$ and $\mathcal{L}_{\text{DA}}$, the transductive-based hybrid DA scheme $\text{DA}_H$ can, in terms of target domain regression performance$\text{NMSE}_{\text{ER}}$ and $\rho$, source domain classification OF1 assurance and stability in GGAP, obtain a reliability performance advantage compared to the inductive-based DT mode.
\vspace{-0.12cm}

\section{Conclusion}
This paper proposes a complex-valued lightweight HybridCVLNet with hybrid task and codeword for seeking the solution to the dataset bias of intra-domain heterogeneous dataset with a jointly regularized optimization and the embedded self-attention complex-valued convolution. Meanwhile, a transductive-based hybrid domain adaptation framework to overcome the data drift of the inter-domain online model fine-tuning with the feature space generative adversarial stimulation of pattern adaptation and the category space distribution alignment is derived. Experimental results show that for dataset bias, the proposed HybridCVLNet can obtain stable generalizability and achieve performance gain over the SOTA feedback scheme under an intra-domain heterogeneous dataset. Moreover, the experiment evaluates the transductive-based hybrid domain adaptation scheme of regression performance and robustness that confounds the data drift over the inductive-based direct transfer learning method under two cross-domain settings.
\vspace{-0.12cm}

\bibliographystyle{IEEEtran}
\bibliography{IEEEBIB}

\end{document}